\documentclass[11pt]{article}
\usepackage{setspace}
\usepackage{latexsym,amsthm,amsmath,amssymb,url}
\usepackage{enumitem}
\usepackage[round]{natbib}
\usepackage{eurosym}

\usepackage{graphics}
\usepackage[usenames]{color}

\usepackage{tikzsymbols}

\theoremstyle{definition}
\newtheorem{lemma}{Lemma}

\theoremstyle{definition}
\newtheorem{definition}{Definition}
\theoremstyle{definition}
\newtheorem{example}{Example}

\theoremstyle{definition}
\newtheorem*{theorem*}{Theorem}
\theoremstyle{definition}
\newtheorem{theorem}{Theorem}

\newtheorem{innercustomlemma}{Lemma}
\newenvironment{customlemma}[1]
  {\renewcommand\theinnercustomlemma{#1}\innercustomlemma}
  {\endinnercustomlemma}

\newtheorem{innercustomthm}{Theorem}
\newenvironment{customthm}[1]
  {\renewcommand\theinnercustomthm{#1}\innercustomthm}
  {\endinnercustomthm}

\newcommand{\red}[1]{{\color{red}#1}}
\newcommand{\blue}[1]{{\color{blue}#1}}

\newcommand{\first}{1^{\text{st}}}
\newcommand{\second}{2^{\text{nd}}}

\newcommand{\set}[1]{\left\{#1\right\}}
\newcommand{\Set}[1]{\Bigl\{#1\Bigr\}}

\newcommand{\Workers}{W}
\newcommand{\worker}[1]{w_{#1}}
\newcommand{\Firms}{F}
\newcommand{\firm}[1]{f_{#1}}

\newcommand{\pref}[1]{\ensuremath{\succ_{#1}}}
\newcommand{\indiff}[1]{\ensuremath{\sim_{#1}}}
\newcommand{\match}{\mu}
\newcommand{\unattractive}{\Chair}
\newcommand{\reduce}{\mathcal{R}}

\usepackage[pdftex, citecolor=blue, colorlinks]{hyperref}

\begin{document}

\title{Unique stable matchings\thanks{We are grateful to Jesper Bagger, Claudia Cerrone, Dongkyu Chang, Vince Crawford, Julie Cullen, Audrey Hu, Sung-Ha Hwang, Miguel Mel\'{e}ndez-Jim\'{e}nez, Kohei Kawaguchi, Christian Keeling, Onur Kesten, Chulyoung Kim, Bettina Klaus, Yunan Li, Wooyoung Lim, Jordi Mass\'{o}, Frances P.\ Ruane, Joel Sobel, Rui Tang, Alex Teytelboym, Alexander Vickery, Joel Watson, Qinggong Wu, Wenji Xu, and participants in seminar presentations at City University of Hong Kong, Hong Kong University of Science and Technology, Korea Advanced Institute of Science and Technology, Universidad de M\'{a}laga, and Yonsei University for helpful comments. Two anonymous referees and an associate editor made many excellent suggestions that improved the paper considerably. Any remaining errors are ours. Anders Yeo's research was supported by the Danish research council for independent research under grant number DFF 7014-00037B.}}
\author{
Gregory Z. Gutin\footnote{Computer Science Department, Royal Holloway University of London.} \hspace{.1in} Philip R.\ Neary\footnote{Economics Department, Royal Holloway University of London.} \hspace{.1in} Anders Yeo\footnote{IMADA, University of Southern Denmark.} $^{,}$\footnote{Department of Mathematics, University of Johannesburg.}
}

\date{\today}

\maketitle

\begin{abstract}
\noindent
In this paper we consider the issue of a unique prediction in one-to-one two-sided matching markets,  as defined by \cite{GaleShapley:1962:AMM}, and we prove the following:
\begin{theorem*}
Let $P$ be a one-to-one two-sided matching market and let $P^{*}$ be its associated normal form, a (weakly) smaller matching market with the same set of stable matchings, that can be obtained using procedures introduced in \cite{IrvingLeather:1986:SIAM} and \cite{BalinskiRatier:1997:}.
The following three statements are equivalent:
\begin{enumerate}[label=(\alph*)]
 \item\label{thmStatementUnique} $P$ has a unique stable matching.
 \item\label{thmStatementAcyclic} Preferences on $P^{*}$ are acyclic, as defined by \cite{Chung:2000:GEB}.
\item\label{thmStatementSingleton} In $P^{*}$ every market participant's preference list is a singleton.
\end{enumerate}
\end{theorem*}
\end{abstract}



\newpage

\section{Introduction}\label{sec:intro}

In this paper we reconsider the classic model of one-to-one two-sided matching, known popularly as the {\it stable matching problem}.
Variations on the framework, introduced first in \cite{GaleShapley:1962:AMM}, have been applied widely to settings ranging from school admissions \citep{AbdulkadirogluSonmez:2003:AER,AbdulkadirogluPathak:2005:AER}, labour markets \citep{CrawfordKnoer:1981:E,KelsoCrawford:1982:E}, the market for kidney donors \citep{RothSonmez:2004:QJE}, and others (see \cite{Roth:2008:IJGT} for a survey).

Yet despite all the attention that the model has received, a very basic question remained unanswered:
\begin{quote}
    {\it Under what conditions is there exactly one stable matching?}
\end{quote}
In this paper we resolve the above open question.
We provide {\it necessary} and {\it sufficient} conditions on the preferences of market participants that guarantee a unique stable matching.

A unique prediction is typically viewed as a desirable property of any economic model since it saves the analyst from an ``equilibrium selection'' headache.
However, as regards two-sided matching markets, determining if there is a unique stable matching is not only of theoretical interest but also of genuine practical importance.
First, there is a literature, beginning with \cite{Roth:1989:JET}, that highlights the role of incomplete information in matching markets and uniqueness plays an important role therein.
Second, unique stable matchings seem to appear disproportionately often in real-world matching markets, and hopefully an understanding of their origin might help explain why.\footnote{Uniqueness has been observed in the National Resident Matching Program \citep{RothPeranson:1999:AER}, Boston school choice \citep{PathakSonmez:2008:AER}, online dating \citep{HitschHortacsu:2010:AER}, and the Indian marriage market \citep{BanerjeeDuflo:2013:AEJMicro}.}
Finally, uniqueness is necessary for the truthful reporting of preferences to be \textit{strategy proof} \citep{GaleSotomayor:1985:AMM, DemangeGale:1987:DAM, Sonmez:1999:E}.\footnote{\cite{GaleSotomayor:1985:AMM} show that if a matching is to be generated by the deferred acceptance algorithm of \cite{GaleShapley:1962:AMM}, then, unless there is a unique stable matching, it is always beneficial for at least one participant to misrepresent their true preferences. Example 1 in \cite{Romero-Medina:2021:IJGT} shows that a unique stable matching is not sufficient for strategy proofness. In incomplete information environments, \cite{EhlersMasso:2007:GEB} show that truth-telling is an ordinal Bayesian Nash equilibrium of the revelation game induced by a common belief and a stable mechanism if and only if all the profiles in the support of the common belief have singleton cores.}
Given this, a better understanding of what assures uniqueness could potentially be levered to ensure both greater efficiency and greater transparency in practice.

While determining {\it if} a matching problem has a unique stable matching is important, determining {\it when} a matching problem has a unique stable matching is actually very easy. Simply run the deferred acceptance algorithm of \cite{GaleShapley:1962:AMM} twice, once with each side in the role of proposer, and check if the stable matching found for each run is the same. If yes, then there is a unique stable matching; if no, then there are at least two. Since the deferred acceptance algorithm runs in polynomial time, this algorithmic approach provides an efficient solution to the uniqueness issue when viewed through a computational lens.
However, while computationally efficient, this algorithmic approach sheds no light on the structure of those matching markets that possess a unique stable matching.
As such, and perhaps also due to the problem's importance, a host of conditions on preferences that are {\it sufficient} for uniqueness have been proposed.\footnote{\label{fn:sufficient}Examples include:\ the {sequential preference condition} \citep{Eeckhout:2000:EL}, the no crossing condition \citep{Clark:2006:BEJTE}, the co-ranking condition \citep{LegrosNewman:2010:EL}, the acyclicity condition \citep{Romero-MedinaTriossi:2013:EL}, the universality condition \citep{HolzmanSamet:2014:GEB}, oriented preferences \citep{Reny:2021:ETB}, and aligned preferences \citep{FerdowsianNiederle:2022:}. The concept of $\alpha$-reducibility \citep{Alcalde:1994:BEJTE,Clark:2006:BEJTE} is both necessary and sufficient for a matching market and any of its submatching markets to have a unique stable matching.}

Perhaps the reason a {\it necessary} {and {\it sufficient}} condition for uniqueness had not been found until now is that seeking conditions on the full preference lists is excessive.
Rather, all that matters for the set of stable matchings is how preferences are defined on the essential part of any matching problem, the subproblem that we term the {\it normal form}.\footnote{The submatching market that is the normal form appeared first in \cite{IrvingLeather:1986:SIAM} and independently in \cite{BalinskiRatier:1997:}. \cite{IrvingLeather:1986:SIAM} did not use a term to describe the normal form. \cite{BalinskiRatier:1997:} used graph-theoretic terminology and referred to it as a {\it domination-free marriage graph}.}
Our main result, Theorem \ref{thm:uniqueStableE}, shows that a unique stable matching is equivalent to preferences being {\it acyclic} \citep{Chung:2000:GEB} on the normal form which is in turn equivalent to the normal form being the unique stable matching and nothing more.
The second equivalence of Theorem~\ref{thm:uniqueStableE} is striking because while the normal form being the unique stable matching is by definition acyclic, the reverse implication is unexpected.

Let us now explain the concepts of acyclic preferences and the normal form.

In a two-sided market with workers on one side of the market and firms on the other,\footnote{While ``workers'' and ``firms'' are the terminology that we employ throughout, we emphasise that these terms are just placeholders. As mentioned in the opening paragraph, the framework is applied widely.} the shortest possible cycle involves two workers and two firms, has length 4, and is described as follows:\ the $\first$ worker prefers the $\second$ firm to the $\first$ firm, the $\second$ firm prefers the $\second$ worker to the $\first$ worker, the $\second$ worker prefers the $\first$ firm to the $\second$ firm, and the $\first$ firm prefers the $\first$ worker to the $\second$ worker.
Acyclic preferences are simply those that do not possess a cycle of this kind.
However, requiring that preferences are acyclic is by itself only sufficient for a unique stable matching \citep{Romero-MedinaTriossi:2013:EL}.
To add necessity, one need only require the weaker condition that preferences are acyclic on the normal form.
In Example \ref{exUnique} we present a matching market with preferences that are not acyclic, and yet there is a unique stable matching since preferences are acyclic on the normal form.

So what then is the normal form of a matching problem? The normal form is arrived at by stripping away parts of individual preference lists that are not relevant to the set of stable matchings.\footnote{Reducing a mathematical object to its bare-boned constituents, the so-called ``canonical form'' or ``normal form'', occurs not only in game theory. It is a common approach in many branches of both pure and applied mathematics (see \href{https://en.wikipedia.org/wiki/Canonical_form}{https://en.wikipedia.org/wiki/Canonical\_form}).}
We term the procedure that discards such irrelevant information the {\em iterated deletion of unattractive alternatives} (IDUA).
The IDUA procedure works by repeatedly pivoting around a particular kind of instability, building on the observation that most preferred partners play a hugely important, albeit somewhat subtle, role.
Suppose that firm $\firm{}$ is worker $\worker{}$'s most preferred place of employment.
Now consider a matching in which $\firm{}$ is paired with a worker that it prefers less than $\worker{}$.
Such a matching cannot be stable, since $\firm{}$ would propose pairing up with $\worker{}$ and $\worker{}$ would certainly accept.
This is because when it comes to top choices, half of the blocking pair is assured.
Worker $\worker{}$ is an outside option for $\firm{}$, and so we say that all the workers less preferred by $\firm{}$ to $\worker{}$ are \emph{unattractive} to $\firm{}$.
{Relating this to the concept of a {\it reservation wage} in labour markets, worker $\worker{}$ is in a sense the current {\it reservation partner} for firm $\firm{}$.}

Whenever $\firm{}$ can guarantee doing better than the unattractive workers in every stable matching, these workers can be deleted from $\firm{}$'s preference list since their presence on $\firm{}$'s preference list is immaterial to the set of stable matchings.
Similarly, all the workers deleted by $\firm{}$ will realise that a (stable) matching with them paired with $\firm{}$ ain't happening and so will delete $\firm{}$ from their preference lists; unattractiveness is reciprocated.
The new matching environment is strictly smaller than the original environment and yet, by definition, the set of stable matchings can not have changed.
But most importantly, further rounds of deletion may now be possible with the shortened preference lists because market participants that were not initially unattractive can become so.
That is, one's outside option / reservation partner can only ever improve.
Eventually the deletion procedure can go no more and what remains we call the normal form.\footnote{This confirms accepted wisdom that the input to a typical matching problem is ``too'' big. That is, more information is supplied than is required. This is consistent with the fact that in modern-day labour markets, in which application costs are close to zero (just one extra click), recruitment resources are heavily weighed down.}

Formally, the IDUA procedure that reduces a matching problem to its normal form parallels closely the iterated deletion of dominated strategies (IDDS) procedure for strategic games (see \cite{Gale:1953:PNASUSA} and \cite{Moulin:1979:E}).
Once IDDS stops, the set of surviving strategies are the only ``rational'' way to play a game in which all players are rational and there is common knowledge of this fact.
That is, IDDS strips a game of strategies that no rational individual who fully understands the environment could ever justify choosing; the resulting game is smaller and yet the set of solutions remains unchanged.
The IDUA procedure has a similar effect on matching problems.
While IDUA discards information, no ``important'' information is discarded in the sense that it only deletes pairs that cannot be part of any stable matching; the resulting matching problem is smaller and yet the set of solutions remains unchanged.\footnote{{So, from a practical perspective, running IDUA should be the first port of call when considering a matching problem for which stability is a requirement. IDUA reduces the size of the input, rendering it more tractable, without affecting the sought-after output (the set of stable matchings). In much the same way, one often begins the analysis of a strategic game by looking for dominated strategies.}} In Section \ref{subsec:IDUA} we compare and contrast these two procedures with particular focus on the sorts of higher-order reasoning required for each to be operationalised.\footnote{Just as an ultra-sophisticated player can determine when a game is dominance solvable, so too could an ultra-sophisticated market participant determine when a matching problem has a unique stable matching.}

While the IDUA procedure can be justified by inferences based on higher ordering reasoning, it also admits another, more practical, interpretation that we believe is interesting.
The deferred acceptance algorithm of \cite{GaleShapley:1962:AMM} is often interpreted as an interactive forum wherein those on one side of the market are assigned as active {proposers}, while those on the other side are relegated to the role of passive {responders}.
In a similar vein, the IDUA procedure can be envisaged as a dynamic marketplace in which, perhaps more realistically, every market participant is simultaneously proposing, responding to proposals, and also handing out preemptive rejections.
This seems to us closer to how two-sided matching markets without a centralised authority operate.\footnote{In a two-sided labour market this would imply that, without knowledge of the normal form, workers apply to too many jobs thereby weighing down hiring resources. Thanks to Julie Cullen, Frances Ruane, Joel Sobel, and Joel Watson for pointing this out.}
While Theorem \ref{thm:uniqueStableE} confirms that a matching market operating in such fashion will not fully ``clear'' unless there is a unique stable matching, the market will invariably become smaller and easier to parse.\footnote{\label{fn:stopIDUA}The market will only not become smaller in the statistically rare and economically unusual case that (i) every worker has a different favourite firm, (ii) every firm has a different favourite worker, and (iii) every market participant's favourite partner views them as least desirable. However, while statistically rare for the market as a whole, Lemma~\ref{lemma:extremalMatches} shows that this is precisely the stopping condition of the IDUA procedure.}
Interpreted in this way, IDUA has the flavour of a bargaining situation or price discovery mechanism typically more associated with traditional ``markets with prices'', in the sense that it closes in on the upper and lower boundaries of the (feasible region of) Pareto efficient allocations.\footnote{For a more classical setting with this kind of feature, consider a seller who owns an object and values it at \euro5, and a buyer who values the same object at \euro10. In a real-world bargaining situation the buyer may start out with a bid short of \euro5 while the seller may first make a demand in excess of \euro10; only when one party crosses into the interval $[$\euro5, \euro10$]$ does the ``real'' bargaining begin. Viewed like this, the deferred acceptance algorithm bestows upon the proposing side in a two-sided market the sort of extreme bargaining power afforded the proposer in the ultimatum game \citep{Harsanyi:1961:JCR,GuthSchmittberger:1982:JEBO}.}

The IDUA procedure generates ever-shortening preference lists and ever-shortening preference lists can be cumbersome to manage.
Fortunately, the directed graph (digraph) approach to matching introduced by \cite{Maffray:1992:JCTB} allows all the information of a matching problem to be encoded on a single digraph instead of multiple preference lists, allowing reductions to be handled in a tractable way.
Maffray's equivalent formulation is easily explained.
Instead of ``lining up'' the participants on opposite sides of the market, one constructs a grid wherein each row corresponds to a specific worker, each column corresponds to a specific firm, and every vertex in the grid corresponds to a pair.
Preferences are depicted by horizontal arcs (directed edges) for workers and vertical arcs for firms.
The unit of measurement in this set up is the pair and not the individual.
A matching is a subset of vertices, pairs, no two of which are in the same row nor the same column.
The acyclic preferences of \cite{Chung:2000:GEB} correspond exactly to the absence of a directed cycle in the matching digraph.

The paper proceeds as follows. Section \ref{sec:Digraph} introduces the matching environment that we consider, the IDUA procedure, and the normal form.\footnote{Many of the properties of IDUA used in the proof of our main result, Theorem \ref{thm:uniqueStableE}, were already obtained by \cite{BalinskiRatier:1997:} using matching digraphs. So that our paper is self-contained, we prove every property that we use and whenever a result of ours is similar to one of \cite{BalinskiRatier:1997:} we explicitly state as such.  Our main result, Theorem \ref{thm:uniqueStableE}, is completely new.}
Section \ref{sec:Results} presents our result classifying uniqueness.
In Section \ref{sec:Digraphs} we introduce the digraph formulation of the matching problem, and using this setup we show how illustrate how the IDUA procedure works.
Section \ref{sec:Conclusion} concludes and discusses potential avenues for future work. 

All proofs appear in Appendix~\ref{app:proofs}.
In Appendix~\ref{app:Equivalent} we show how running IDUA ``in reverse'' allows an analyst to identify the full collection of matching markets for which a given matching is the unique stable matching.

\section{Matching problems and their normal form}\label{sec:Digraph}

In Section \ref{subsec:environment} we introduce the one-to-one matching environment with complete preference lists.
In Section \ref{subsec:unattractiveness} we define what it means for market participants to be deemed unattractive.
In Section \ref{subsec:IDUA} we introduce the IDUA procedure that repeatedly prunes information from preference lists that is irrelevant to the set of stable matchings of a given matching market.
When the IDUA procedure stops, the resulting environment is termed the normal form.

\subsection{Matching problems}\label{subsec:environment}

Let $\Workers$ be a set of $n$ workers and let $\Firms$ be a set of $n$ firms.
(We fix $n$ as a positive integer greater than or equal to 2.)
Each worker $\worker{} \in \Workers$ has a strict preference relation, $\pref{\worker{}}$, over the set of firms, and each firm $\firm{} \in \Firms$ has a strict preference relation, $\pref{\firm{}}$, over the set of workers.
When worker $\worker{}$ prefers firm $\firm{}'$ to firm $\firm{}''$, we will write $\firm{}' \pref{\worker{}} \firm{}''$, with an analogous statement for the preferences of firms.
A preference relation $\pref{\worker{}}$ is said to be {\em complete} if all firms are in the relation; similarly for $\pref{\firm{}}$.
We assume throughout that all preference relations are complete.\footnote{While we have defined a preference relation as a binary relation over a set of participants, occasionally we will view preferences as ordered lists where the first entry on the list is the most preferred participant, and so on.}

The following is the environment that we consider in this paper.
It is precisely the environment originally considered by \cite{GaleShapley:1962:AMM}. 

\begin{definition}\label{def:instance}
An instance of the stable matching problem, $P$, is defined as the pair $(\set{\pref{\worker{}}}_{\worker{}\in \Workers}, \set{\pref{\firm{}}}_{\firm{}\in \Firms})$, where $\set{\pref{\worker{}}}_{\worker{}\in \Workers}$ and $\set{\pref{\firm{}}}_{\firm{}\in \Firms}$ are the collection of complete preference relations, one for each worker and firm.
\end{definition}

A \emph{matching} in $P$ is a mapping $\match$ from $\Workers \cup \Firms$ to itself such that:\ for every worker $\worker{} \in \Workers$, $\match(\worker{}) \in \Firms$; for every firm $\firm{} \in \Firms$, $\match(\firm{}) \in \Workers$; and for every $\worker{}, \firm{} \in \Workers \cup \Firms$, $\match(\worker{}) = \firm{}$ if and only if $\match(\firm{}) = \worker{}$.

The following is the key definition proposed by \cite{GaleShapley:1962:AMM}.

\begin{definition}\label{def:blocking}
Worker $\worker{}$ and firm $\firm{}$ form a \textit{blocking pair} with respect to matching $\match$ in $P$, if $\firm{} \pref{\worker{}} \match(\worker{})$ and $\worker{} \pref{\firm{}} \match(\firm{})$.
\end{definition}

In words, $(\worker{}, \firm{})$ form a blocking pair with respect to matching $\match$ if both $\worker{}$ and $\firm{}$ prefer each other over their partners in $\match$.
That is, $\worker{}$ and $\firm{}$ would prefer to break away from their current partners and pair up together.
The notion of stability is defined by the absence of a blocking pair.

\begin{definition}\label{def:stableMatch}
A matching $\match$ in $P$ with no blocking pairs is a \textit{stable matching}.
\end{definition}

\cite{GaleShapley:1962:AMM} introduced the now-classic ``deferred acceptance'' algorithm and used it to prove the following existence result.

\begin{theorem*}[\cite{GaleShapley:1962:AMM}]\label{thm:GaleShapley}
Every instance of the stable matching problem possesses at least one stable matching.
\end{theorem*}

The following example illustrates the above definitions.

\begin{example}\label{exInitial}
Let $P_{1}$ be an instance of the stable matching problem in which there are three workers and three firms.
Precisely, $\Workers = \set{\worker{1}, \worker{2}, \worker{3}}$ and $\Firms = \set{\firm{1}, \firm{2}, \firm{3}}$, and preferences are as follows.\begin{align*}
\worker{1}: \firm{2} \pref{\worker{1}} \firm{1} \pref{\worker{1}} \firm{3} & \hspace{.4in}  \firm{1}: 	\worker{1} \pref{\firm{1}} \worker{2} \pref{\firm{1}} \worker{3} 			\\
\worker{2}: \firm{2} \pref{\worker{2}} \firm{3} \pref{\worker{2}} \firm{1}  & \hspace{.4in}  \firm{2}: 	\worker{1} \pref{\firm{2}} \worker{2} \pref{\firm{2}} \worker{3}			\\
\worker{3}: \firm{1} \pref{\worker{3}} \firm{2} \pref{\worker{3}} \firm{3}  & \hspace{.4in}  \firm{3}: 	\worker{1} \pref{\firm{3}} \worker{3} \pref{\firm{3}} \worker{2} 
\end{align*}

It can be checked that $P_{1}$ possesses two stable matchings, that we label $\match_{1}$ and $\match_{2}$. They are,
\begin{equation}\label{eq:example1stable}
\begin{aligned}
\match_{1}:  \hspace{.3in} (\worker{1}, \firm{2}), (\worker{2}, \firm{3}), (\worker{3}, \firm{1}) \\
\match_{2}:  \hspace{.3in} (\worker{1}, \firm{2}), (\worker{2}, \firm{1}), (\worker{3}, \firm{3})
\end{aligned}
\end{equation}

\end{example}

Example \ref{exInitial} is straightforward.
We introduce it because in the next section we use it to highlight how an instance of the stable matching problem may contain more information than is required in order to compute the full set of stable matchings for a particular two-sided market.

To give a taster of what we mean by the above, let us briefly consider Example \ref{exInitial} from the perspective of worker $\worker{1}$ and from the perspective of firm $\firm{2}$.
Since $\worker{1}$'s most preferred firm is $\firm{2}$ and $\firm{2}$'s most preferred worker is $\worker{1}$, it must be that the pair $(\worker{1}, \firm{2})$ are in all stable matchings since $(\worker{1}, \firm{2})$ would form a blocking pair against any matching not including it.
(This is corroborated by the two stable matchings of Example \ref{eq:example1stable}, $\match_{1}$ and $\match_{2}$, given in \eqref{eq:example1stable}.)

The above observation can be built on.
Given that $\worker{1}$ will certainly be matched with (their favourite firm) $\firm{2}$ in every stable matching, the fact that $\worker{1}$ has a relative preference for $\firm{1}$ over $\firm{3}$ is irrelevant.
By this we mean the following:\ observe that if $\worker{1}$'s preferences were altered so that their relative preference for $\firm{1}$ over $\firm{3}$ were swapped, the set of stable matchings would remain unchanged.
In fact, it is also the case that if worker $\worker{1}$'s preferences were incomplete, such that they preferred to be unmatched over being matched with either $\firm{1}$ or $\firm{3}$, the set of stable matchings would further remain unchanged.
This motivates the notion of an {\it unattractive alternative} that is the subject of the next section.

\subsection{Unattractive alternatives}\label{subsec:unattractiveness}

The informal discussion following Example~\ref{exInitial} above highlights that for some instances of the matching problem the input supplied may exceed that which is required to compute the set of stable matchings.
That is, since worker $\worker{1}$ in Example \ref{exInitial} is so highly sought-after, the environment would be the same were $\worker{1}$ to have incomplete preferences.
But then the position in which worker $\worker{1}$ appears on the preference lists of the firms with whom $\worker{1}$ will not match is also irrelevant to the set of stable matchings.

We formalise this using the symbol $\indiff{}$ to denote indifference and using the symbol $\unattractive$ as a placeholder that is ``as bad'' as unattractive alternatives.

\begin{definition}[Unattractive alternatives]\label{def:unattractive}
We say that 
\begin{enumerate}[label=(\roman*)]
\item\label{unattractive1}
firm $\firm{}$ is an {unattractive alternative} to worker $\worker{}$, denoted $\unattractive \indiff{\worker{}} \firm{}$, if there is some firm $\firm{}' \neq \firm{}$ such that (i) $\firm{}' \pref{\worker{}} \firm{}$, and (ii) $\worker{} \pref{\firm{}'} \worker{}'$ for all $\worker{}' \neq \worker{}$.
\item\label{unattractive2}
worker $\worker{}$ is an {unattractive alternative} to firm $\firm{}$, denoted $\unattractive \indiff{\firm{}} \worker{}$, if there is some worker $\worker{}' \neq \worker{}$ such that (i) $\worker{}' \pref{\firm{}} \worker{}$, and (ii) $\firm{} \pref{\worker{}'} \firm{}'$ for all $\firm{}' \neq \firm{}$.
\item\label{unattractive3}
$\firm{}$ is an {unattractive alternative} to worker $\worker{}$ whenever $\worker{}$ is an {unattractive alternative} to firm $\firm{}$ (and vice versa).
\end{enumerate}
\end{definition}

In words, condition \ref{unattractive1} of Definition~\ref{def:unattractive} says the following:\ worker $\worker{}$ deems firm $\firm{}$ unattractive if $\worker{}$ will certainly do better than $\firm{}$ in every stable matching.
This is guaranteed when $\worker{}$ is the most preferred worker of some firm $\firm{}'$ that $\worker{}$ prefers to $\firm{}$, because then $\worker{}$ and $\firm{}'$ would form a blocking pair to any matching that matches $\worker{}$ with $\firm{}$.
Condition \ref{unattractive2} of Definition~\ref{def:unattractive} is the analog to condition \ref{unattractive1} but for firms instead of workers. Condition \ref{unattractive3} stipulates that unattractiveness is reciprocated.

We pause briefly to note that there is an existing notion in the literature of an {\it unacceptable alternative}.
Worker $w$ views firm $f$ as unacceptable if $w$ would prefer to be unmatched over being paired with $f$.
Unacceptable alternatives go hand in hand with preferences not being complete.
Our notion of an unattractive alternative, on the other hand, is compatible with preferences being complete.
In fact, as we will now see, some unattractive alternatives can be extremely desirable but are deemed a``waste of time'' given the structure of the market.

Let us now revisit Example~\ref{exInitial} using the concept of unattractive alternatives. 
We begin by considering condition \ref{unattractive1}.
As noted previously, worker $\worker{1}$ is firm $\firm{2}$'s most preferred worker.
Given this the other two firms $\firm{1}$ and $\firm{3}$ are unattractive to $\worker{1}$.
It then follows that $\firm{2} \pref{\worker{1}} \unattractive \indiff{\worker{1}} \set{\firm{1}, \firm{3}}$, where we have gathered the collection of $\worker{1}$'s unattractive alternatives in a set in which the order that they are listed is immaterial.
By a similar reasoning, condition \ref{unattractive2} yields $\worker{1} \pref{\firm{2}} \unattractive \indiff{\firm{2}} \set{\worker{2}, \worker{3}}$.

Consider now how the above statements impact instance $P_{1}$ of Example~\ref{exInitial}.
It is clear that $P_{1}$ is, from the perspective of stability, identical to the instance $P_{1}'$, where $P_{1}'$ is defined as, $\Workers = \set{\worker{1}, \worker{2}, \worker{3}}$ and $\Firms = \set{\firm{1}, \firm{2}, \firm{3}}$, and preferences are as follows:\footnote{To emphasise that instance $P_{1}$ is equivalent to this smaller instance, $P_{1}'$, we omit from a participant's preference list those that are equivalent to $\unattractive$.}
\begin{align*}
&\worker{1}: \firm{2} \pref{\worker{1}} \unattractive & \hspace{.4in} & \firm{1}:  \worker{1} \pref{\firm{1}} \worker{2} \pref{\firm{1}} \worker{3} \pref{\firm{1}} \unattractive 	\\
&\worker{2}: \firm{2} \pref{\worker{2}} \firm{3} \pref{\worker{2}} \firm{1} \pref{\worker{2}} \unattractive  & \hspace{.4in} & \firm{2}: 	\worker{1} \pref{\firm{2}} \unattractive 	\\
&\worker{3}: \firm{1} \pref{\worker{3}} \firm{2} \pref{\worker{3}} \firm{3} \pref{\worker{3}} \unattractive &  \hspace{.4in} & \firm{3}: 	\worker{1} \pref{\firm{3}} \worker{3} \pref{\firm{3}} \worker{2} \pref{\firm{3}} \unattractive 
\end{align*}

To illustrate condition \ref{unattractive3} from Definition \ref{def:unattractive} consider the following.
Since worker $\worker{1}$ has deemed both firms $\firm{1}$ and $\firm{3}$ unattractive, condition \ref{unattractive3} requires that both $\firm{1}$ and $\firm{3}$ reciprocate. The reason for this is that a match with $\worker{1}$ is not happening for either of these firms, so maintaining $\worker{1}$ in one's preference list serves no purpose.
In particular we note that firm $\firm{1}$ deems $\worker{1}$ as unattractive despite the fact that $\firm{1}$ appears first in $\worker{1}$'s preference list.
A similar statement holds for workers $\worker{2}$ and $\worker{3}$, both of whom reciprocate unattractiveness to firm $\firm{2}$. This means that instance $P_{1}'$ above is also identical to instance $P_{1}''$, where $P_{1}''$ is defined as, $\Workers = \set{\worker{1}, \worker{2}, \worker{3}}$ and $\Firms = \set{\firm{1}, \firm{2}, \firm{3}}$, and preferences are as follows:
\begin{align*}
& \worker{1}: \firm{2} \pref{\worker{1}} \unattractive  & \hspace{.4in} & \firm{1}:  \worker{2} \pref{\firm{1}} \worker{3} \pref{\firm{1}} \unattractive 	\\
& \worker{2}: \firm{3} \pref{\worker{2}} \firm{1} \pref{\worker{2}} \unattractive & \hspace{.4in} & \firm{2}: 	\worker{1} \pref{\firm{2}} \unattractive 	\\
& \worker{3}: \firm{1} \pref{\worker{3}} \firm{3} \pref{\worker{3}} \unattractive & \hspace{.4in} & \firm{3}: 	\worker{3} \pref{\firm{3}} \worker{2} \pref{\firm{3}} \unattractive 
\end{align*}

Let us now make some observations.
The first, and it is easily verified, is that the set of stable matchings for instance $P_{1}''$ coincides precisely with that of $P_{1}$.
The second is that the set of stable matchings coincide despite the fact that $P_{1}''$ is, in a precise sense, strictly smaller than instance $P_{1}$.
To see this we note two features:\ (i) in $P_{1}''$, every market participants' preference list is no longer than in $P_{1}$ (in fact each is strictly shorter), and (ii) the relative ordering of any pair in a preference list of $P_{1}''$ is the same as for $P_{1}$.
Instance $P_{1}''$ contains all the relevant information of $P_{1}$ and yet is simpler to parse.

The above hints that (mutually) unattractive alternatives play no role in the set of stable matchings of any instance of the matching problem.
In many ways, unattractive alternatives have much the same effect on stable matchings as strictly dominated strategies have on strategic games. We recall that deleting strictly dominated strategies from a strategic environment reduces the size of the game and yet does not change the set of rationalisable outcomes nor the set of equilibria (the set of predictions).
One might wonder whether deleting unattractive alternatives has the effect of reducing the input to a matching problem and yet does not affect the set of stable matchings (the set of predictions).
The answer turns out to be yes. 

Another natural question to ask is whether deleting unattractive alternatives from an instance of the matching problem can only be performed once. When deleting strictly dominated strategies, the deletion operation can be applied on the reduced version of a game since strategies that were not initially strictly dominated can become so. The same occurs with unattractive alternatives, and the reason is that there may be participants who were not initially unattractive but are unattractive in the reduced environment. We address this in the next section wherein we define the {\it iterated deletion of unattractive alternatives} (IDUA), a procedure that parallels closely the iterated deletion of dominated strategies (IDDS) due to \cite{Gale:1953:PNASUSA} and applied to voting by \cite{Moulin:1979:E}.

\subsection{IDUA and a matching problem's normal form}\label{subsec:IDUA}

We now define the  {\em iterated deletion of unattractive alternatives} (IDUA), a procedure that repeatedly prunes redundant information from the preference lists.
Technically it continually deletes unattractive participants from preference lists until there remains no market participant who views any other as unattractive.

\begin{definition}[The iterated deletion of unattractive alternatives (IDUA)]\label{def:IDUAe}
Given an instance of the matching problem $P = \big(\set{\pref{\worker{}}}_{\worker{}\in \Workers}, \set{\pref{\firm{}}}_{\firm{}\in \Firms}\big)$, we define $\pref{\worker{}}^{0} \, := \, \pref{\worker{}}$ and $\pref{\firm{}}^{0} \, := \, \pref{\firm{}}$, and for each $k \geq 1$, form the matching (sub)problem $P^{k} = \big(\{\pref{\worker{}}^{k}\}_{\worker{}\in \Workers}, \{ \pref{\firm{}}^{k}\}_{\firm{}\in \Firms } \big)$ where for every worker $\worker{}$ and every firm $\firm{}$,
\begin{equation}\label{eq:iterative}
\begin{aligned}
\pref{\worker{}}^{k} &= \set{\firm{}  \, | \,\, \firm{} \pref{\worker{}}^{k-1} \unattractive \, \text{ and } \, \worker{} \pref{\firm{}}^{k-1} \unattractive }, \text{ and }\\
\pref{\firm{}}^{k} &= \set{\worker{} \,  | \,\, \worker{} \pref{\firm{}}^{k-1} \unattractive \, \text{ and } \, \firm{} \pref{\worker{}}^{k-1} \unattractive}.
\end{aligned}
\end{equation}
Finally, define the {\em normal form} of matching problem $P$, $P^{*}$, as $P^{k^{*}}$ where $k^*$ is the minimum $k$ such that $P^{k+1} = P^{k}$. That is, the normal form $P^{*}$ is what remains when no further deletions are possible for some $k$.
Worker preferences on the normal form are denoted $\set{\pref{\worker{}}^{*}}$ and firm preferences by $\{\pref{\firm{}}^{*}\}$.
\end{definition}
The iterative part of Definition \ref{def:IDUAe}, given in \eqref{eq:iterative}, says that if worker $\worker{}$ and firm $\firm{}$ do not find each other mutually unattractive at some round of the iteration procedure, then neither deletes the other from their preference list during that round. That is, worker $\worker{}$ carries firm $\firm{}$ forward to the next round of the procedure and vice versa.

The IDUA procedure has parallels with the IDDS procedure for strategic games that are quite striking. Once the IDDS procedure stops (and it must), the set of surviving strategies are the only ``rational'' way to play the game. The following result, whose proof is in the Appendix, confirms a similar feature of the IDUA procedure. Precisely, it shows that while the IDUA procedure may discard information from a matching problem, no ``important'' information is discarded in the sense that the set of stable matchings for $P$ can be computed using only $P^{*}$.
That is, preference lists are pruned in such a way that the set of stable matchings remains unchanged.\footnote{\cite{IrvingLeather:1986:SIAM} introduce an alternative deletion procedure that is discussed in detail in Section 3.2.1 of \cite{RothSotomayor:1990:}. While the procedure is purely mechanical, in that it does not have a behavioural interpretation like IDUA, using the second part of Lemma~\ref{lemma:orderIndependentE} it can be shown that this procedure also reduces a matching problem to its normal form.}

\begin{lemma}[\cite{BalinskiRatier:1997:}] \label{lemma:iduaSameMatchingsE}
The iterated deletion of unattractive alternatives does not change the set of stable matchings.  That is, $P$ and its normal form, $P^{*}$, contain exactly the same set of stable matchings.
\end{lemma}

Let us now make some observations.
The first concerns the details of the IDUA procedure that arrives at the normal form.
In the discussion of unattractive alternatives following Example \ref{exInitial}, and also in the formal statements of Definition~\ref{def:IDUAe}, statements about deleting unattractive alternatives were made in a particular order.
A natural concern then is whether the order in which the unattractive alternatives are deleted might matter.
Fortunately, the first part of Lemma~\ref{lemma:orderIndependentE} below (that also appeared in \cite{BalinskiRatier:1997:}), whose proof is in the Appendix, shows that there is no issue with this as the resulting normal form is arrived at independently of the order in which unattractive alternatives are removed.\footnote{An analogous result holds for IDDS when applied to finite games (see \cite{Mas-ColellWhinston:1995:} Exercise 8.B.4), though care must be taken with infinite games (see \cite{DufwenbergStegeman:2002:E}).}

Concerning the mathematical structure of the normal form, one might think that the following conjecture should be true (but it turns out not to be):\ if worker $\worker{}$ and firm $\firm{}$ do not find each other unattractive at any point (i.e., $\firm{} \in \set{\pref{\worker{}}^{*}}$ and $\worker{} \in \{\pref{\firm{}}^{*}\}$), then the pair $(\worker{}, \firm{})$ is contained in {\it some} stable matching.
The second part of Lemma~\ref{lemma:orderIndependentE} shows that the conjecture is false.
However, it is the case that if the pair $(\worker{}, \firm{})$ is contained in the normal form but is not part of any stable matching, then it must satisfy structural property ($s$).
{When we translate the environment to that of a directed graph in Section~\ref{sec:Digraphs}, property ($s$) will be clearly interpreted as that such a pair must be ``surrounded'' by other pairs that are contained in some stable matching.}

\begin{lemma}[] \label{lemma:orderIndependentE}
Let $P$ be an instance of the matching problem.
Then the normal form of $P$, $P^{*}$, is uniquely defined.
That is, no matter in which order we repeatedly delete unattractive alternatives from preference lists, we always end up with the same $P^{*}$.

Furthermore, suppose that $\firm{} \in \set{\pref{\worker{}}^{*}}$ and $\worker{} \in \{\pref{\firm{}}^{*}\}$.
Then, either $(\worker{}, \firm{})$ is part of some stable matching or the following property ($s$) holds.
\begin{description}
\item[($s$):] There exist firms $\firm{j_1}$ and $\firm{j_2}$ such that $\firm{j_1}, \firm{j_2} \in \{\pref{\worker{}}^{*}\}$ with $\firm{j_1} \pref{\worker{}}^{*} \firm{} \pref{\worker{}}^{*} \firm{j_2}$, and there exist workers $\worker{i_1}$ and $\worker{i_2}$ such that $\worker{i_1}, \worker{i_2} \in \{\pref{\firm{}}^{*}\}$ with $\worker{i_1} \pref{\firm{}}^{*} \worker{} \pref{\firm{}}^{*} \worker{i_2}$.
\end{description}
\end{lemma}

Given that strictly smaller instances are by definition computationally easier to handle, together Lemma \ref{lemma:iduaSameMatchingsE} and Lemma \ref{lemma:orderIndependentE} imply that performing the IDUA procedure should be the first port of call for an analyst who insists on stability in a given two-sided matching market. 

We now document some further connections between the IDDS procedure and the IDUA procedure. The first point to note concerns effectiveness.
In many strategic games, IDDS does not simplify the environment.
IDUA on the other hand almost always has some bite; except in the rare cases that each worker has a different favourite firm, every firm has a different favourite worker, and every market participant's favourite partner ranks them last.

In fact, some reflection reveals that this statement above concerning when IDUA has no impact must also be the stopping condition for the procedure.
Lemma~\ref{lemma:extremalMatchesE} below confirms this.
Before stating the Lemma, we introduce some notation.
Given a matching problem $P = (\set{\pref{\worker{}}}_{\worker{}\in \Workers}, \set{\pref{\firm{}}}_{\firm{}\in \Firms})$, for every worker $\worker{}$, let $\tau(\pref{\worker{}})$ denote the firm first in $\worker{}$'s preference list and let $\tau(\pref{\firm{}})$ denote the first worker in $\firm{}$'s preference list.

\begin{lemma}[\cite{BalinskiRatier:1997:}]\label{lemma:extremalMatchesE}
Let $P$ be an instance of the matching problem and let $P^*$ denote the normal form of $P$.
The following two collection of pairs, $\match_{\Workers}$ and $\match_{\Firms}$, are both stable matchings in $P$.
\begin{center}
$\begin{array}{rcl} \vspace{0.2cm}
 \match_{\Workers} & = & \Set{\big(\worker{1}, \tau(\pref{\worker{1}^{*}})\big), \dots, \big(\worker{n}, \tau(\pref{\worker{n}^{*}})\big)} \\
 \match_{\Firms} & = & \Set{\big(\tau(\pref{\firm{1}^{*}}), \firm{1}\big), \dots, \big(\tau(\pref{\firm{n}^{*}}), \firm{n}\big)} \\
\end{array}$
\end{center}
\end{lemma}

Consider the collection of pairs $\match_{\Workers}$ above.
Every worker is paired with their most preferred firm in the normal form.
For this to be a matching, it must be that every worker has a different favourite firm in the normal form.
This has to be the case because whenever two workers have the same favourite firm, the IDUA procedure cannot be finished.
To see why, suppose that workers $\worker{1}$ and $\worker{2}$ have the same favourite firm $f$, so that $\tau(\pref{\worker{1}^{*}}) = \tau(\pref{\worker{2}^{*}}) = \firm{}$.
Given $\firm{}$ has strict preferences, $\firm{}$ must strictly prefer one of the two workers and will deem the less preferred of them as unattractive.
To see that $\match_{\Workers}$ is not only a matching but also a stable one, we observe that every worker is paired with their most preferred feasible partner, and so there cannot be any blocking pairs since every worker is as content as can be.
Our goal is to classify uniqueness using primitives of the model, but we note, without invoking, that the two stable matchings above are those found by the deferred acceptance algorithm.

Let us now consider how introspective market participants might view a matching market.
Like IDDS and the solution concept of {\it rationalizability} \citep{Bernheim:1984:E,Pearce:1984:E,TanWerlang:1988:JET}, IDUA can be justified by appealing to a form of ``higher order reasoning'' in recognising how other participants view the environment. However, the higher order reasoning invoked is different for IDUA. Both IDDS and rationalizability work by assuming ``rationality'' and ``sophistication'' on the part of individuals:\ rational individuals avoid dominated strategies and sophisticated individuals expect their rational opponents to do the same. And so on. With IDUA there is a ``moment'' at which both participants simultaneously recognise each other's unattractiveness and delete each other from their preference lists. It is then required that third parties are capable of recognising this and processing it.
Third parties do so as their relative placing in the preference lists of others can have changed. While this might seem implausible at first, the solution concept of stability is coalitional in nature, so perhaps it is not unreasonable that the sort of higher order reasoning in which participants engage should be too.

Another way to highlight how the higher order reasoning differs between IDUA and IDDS can be seen by considering how each procedure works in the first round of deletion.
In a strategic game, a player can delete their own strictly dominated strategies without any knowledge of the other players' payoffs.
The same is not true of IDUA in matching markets.
Worker $w$ can only decide that firm $\firm{j}$ is unattractive if there is some other firm, say $\firm{k}$, who worker $w$ prefers to firm $\firm{j}$ and for whom worker $w$ is the most preferred worker.
But for worker $w$ to be able to do this, in addition to knowing their own preferences, importantly worker $w$ must also have knowledge of $\firm{k}$'s preferences.
(We note however that worker $w$ need not know firm $\firm{j}$'s preferences.)

We conclude this section with an observation.
If any two instances of the matching problem possess the same normal form, then they must have the same set of stable matchings.
But one can show by example that the reverse implication does not hold.
That is, it need not be the case that two instances with the same set of stable matchings have the same normal form.
However, the reverse implication does hold for instances that possess the same unique stable matching.
This can be exploited as follows.
Given a matching one can generate all instances of the matching problem for which that matching is the unique stable matching:\ simply start out with the matching in question and run all possible variants of the IDUA procedure ``in reverse''.
Such a procedure is sketched in Appendix~\ref{app:Equivalent}.

\section{Classifying unique stable matchings}\label{sec:Results}

Checking whether an instance of the matching problem has a unique stable matching can be done as follows.
Run the deferred acceptance algorithm of \cite{GaleShapley:1962:AMM} twice, once with workers in the role of proposers and once with firms in the role of proposers, and check if the stable matching found for each run is the same.
If yes, then there is a unique stable matching.
If no, then there are at least two.
The reason for this is that the set of stable matchings form a distributive lattice of which the worker-proposing stable matching and the firm-proposing stable matching are the extreme elements \citep{Knuth:1996:}.
Since the deferred acceptance algorithm runs in polynomial time, the algorithmic approach provides an efficient solution from the perspective of computational complexity.

But while computationally efficient, this purely algorithmic approach sheds no light on the structure of instances that possess a unique stable matching.
It is for this reason that a host of sufficient conditions on preferences ensuring uniqueness have been proposed (see the references in Footnote \ref{fn:sufficient}).
Perhaps the reason a necessary and sufficient condition had not been found before now was that, as per Lemma \ref{lemma:iduaSameMatchingsE}, what really matters is how preferences operate on the normal form.
It turns out that the barrier to uniqueness is preference lists that possess {\it cycles} \citep{Chung:2000:GEB} on the normal form.

\begin{definition}\label{def:cycle}
Let $P = (\set{\pref{\worker{}}}_{\worker{}\in \Workers}, \set{\pref{\firm{}}}_{\firm{}\in \Firms})$ be an instance of the matching problem with $n$ workers and $n$ firms. We say that the preference lists of $P$ possess a {\it cycle}, if there exists a subset of workers of size $k$ and a subset of firms of size $k$ (with $2 \le k \leq n$), and an enumeration of and ordering of the participants $\set{\firm{1}, \worker{1}, \firm{2}, \worker{2}, \firm{3}, \dots, \worker{k-1}, \firm{k}, \worker{k}}$ such that
\begin{equation}\label{eq:cycleDef}
\begin{aligned}
\firm{j+1} &\pref{\worker{j}} \firm{j} &\hspace{.3in} &\text{ for all } j = 1, \dots, k \text{ (modulo } k), \text{ and} \\
\worker{j} &\pref{\firm{j}} \worker{j-1} & \hspace{.3in} &\text{ for all } j = 1, \dots, k \text{ (modulo } k)
\end{aligned}
\end{equation}
We say that an instance $P$ of the matching problem is {\it acyclic} if its preference lists do not possess a cycle.\footnote{In the context of many-to-one matching environments both \cite{Ergin:2002:E} and \cite{Kesten:2006:JET} provide alternative definitions of acyclic preference lists.}
\end{definition}

Each participant's preference list is generated by a binary relation that is antisymmetric and negatively transitive. Together these imply transitivity that effectively translates as ``individually acyclic''. But while each individual preference relation is acyclic, cycles can materialise in the system as a whole due to the interconnectedness of the market.
We defer a detailed discussion of this until Section \ref{sec:Digraphs} because matching digraphs allow the representation of cycles in an intuitive way.

We now state our main result, the proof of which is found in Appendix~\ref{app:proofs}.

\begin{theorem} \label{thm:uniqueStableE}
Let $P$ be an instance of the matching problem and let $P^{*}$ be its associated normal form.
The following three statements are equivalent.
\begin{enumerate}[label=(\alph*)]
 \item\label{thmStatementUnique} $P$ has a unique stable matching.
 \item\label{thmStatementAcyclic} The normal form of $P$, $P^{*}$, is acyclic.
 \item\label{thmStatementSingleton} In the normal form, $P^{*}$, every market participant's preference list is a singleton.
\end{enumerate}
\end{theorem}

The equivalence of \ref{thmStatementUnique} and \ref{thmStatementAcyclic} confirms that it is cycles in preferences on the normal form that prevent uniqueness. To see why, consider a stable matching, $\match$, and consider a subset of market participants, $S$, whose preferences possess a cycle.
Well it turns out that we can then ``shuffle around'' some participants in $S$, by assigning them different partners, also in $S$, and arrive at another stable matching.
To illustrate this let us return again to Example~\ref{exInitial} that possessed two stable matchings. We restrict attention to $P_{1}''$ that is the normal form of $P_{1}$ (in that no further deletions are possible). 

Let $S$ be the subset of market participants $\set{\worker{2}, \worker{3}, \firm{1}, \firm{3}}$.
Restricted to this subset, there are two stable (sub)matchings. They are $(\worker{2}, \firm{1}), (\worker{3}, \firm{3})$ and $(\worker{3}, \firm{1}), (\worker{2}, \firm{3})$. Note that the first of these stable (sub)matchings is firm-optimal, as evidenced by the fact that both firms are with their most preferred partner available in the normal form (while worker $\worker{1}$ is the favourite worker of all firms in the original instance, $P_{1}$, worker $\worker{1}$ was deleted from the preference lists of all firms bar firm $\firm{2}$). Likewise, the second stable (sub)matching is worker-optimal.

When we ``zoom in'' further on the subset of market participants $S$ in the normal form $P_{1}''$, we note that there is a cyclic structure to their collective preferences.
To illustrate this, let us construct a sequence that begins with an arbitrarily chosen participant from this subset, and every subsequent element in the sequence is the most preferred participant of the participant listed before.
As an example, if we begin the sequence with $\worker{2}$, then the sequence is $(\worker{2}, \firm{3}, \worker{3}, \firm{1}, \worker{2}, \dots)$, where the ``\dots'' indicate that the cycle has restarted. This can be formally stated as follows.
\begin{equation}\label{eq:cycle}
\firm{3} \pref{\worker{2}} \firm{1} \text{ and } \worker{3} \pref{\firm{3}} \worker{2} \text{ and } \firm{1} \pref{\worker{3}} \firm{3} \text{ and } \worker{2} \pref{\firm{1}} \worker{3}
\end{equation}
Now if we relabel $\worker{2}$ by $\worker{1'}$,  $\firm{1}$ by $\firm{1'}$,  $\worker{3}$ by $\worker{2'}$, and $\firm{3}$ by $\firm{2'}$, then the expressions in \eqref{eq:cycle} read as
\begin{equation}\label{eq:cycleOrdered}
\firm{2'} \pref{\worker{1'}} \firm{1'} \text{ and } \worker{2'} \pref{\firm{2'}} \worker{1'} \text{ and } \firm{1'} \pref{\worker{2'}} \firm{2'} \text{ and } \worker{1'} \pref{\firm{1'}} \worker{2'}
\end{equation}
where we note that the expressions in \eqref{eq:cycleOrdered} provide an example, with $k=2$, of the condition for a cycle from Definition \ref{def:cycle}.

\cite{Romero-MedinaTriossi:2013:EL} showed that acyclic preferences are sufficient for uniqueness.
From the fact that \ref{thmStatementAcyclic} implies \ref{thmStatementUnique} we can see why this is true:\ if the preference lists are acyclic to begin with, then clearly pruning the preference lists cannot generate a cycle that was not present before.
That is, if the preferences are acyclic to begin with (and hence there is a unique stable matching), then the normal form, that is by definition a (weakly) smaller matching market, must be acyclic too (and hence must possess the same unique stable matching).

Let us now consider the equivalence of parts \ref{thmStatementAcyclic} and \ref{thmStatementSingleton}.
On first inspection \ref{thmStatementSingleton} appears a far stronger condition than \ref{thmStatementAcyclic}, in that clearly \ref{thmStatementSingleton} implies \ref{thmStatementAcyclic}:\ since for preference lists to possess a cycle at least four participants must each have a preference list of length at least 2. 
That \ref{thmStatementAcyclic} implies \ref{thmStatementSingleton} means that once an acyclic (sub)matching problem is reached, the IDUA procedure will continue to truncate the problem until it reduces to the unique stable matching and nothing more.\footnote{Given the similarity between IDDS for strategic games and IDUA for matching problems, matching markets with exactly one stable matching are in a sense the analog of strategic games that are {\it dominance solvable}.}

\begin{figure}[hbtp]
\begin{center}
\tikzstyle{vertexDOT}=[scale=0.25,circle,draw,fill]
\tikzstyle{vertexY}=[circle,draw, top color=gray!5, bottom color=gray!30, minimum size=11pt, scale=0.6, inner sep=0.99pt]
\tikzstyle{vertexW}=[circle,dotted, draw, top color=gray!1, bottom color=gray!1, minimum size=11pt, scale=0.55, inner sep=0.99pt]
\begin{tikzpicture}[scale=1.0]
\node (x11) at (1,5) [vertexW] {$(1,1)$};
\node (x12) at (3,5) [vertexY] {$(1,2)$};
\node (x13) at (5,5) [vertexW] {$(1,3)$};
\node (x21) at (1,3) [vertexY] {$(2,1)$};
\node (x22) at (3,3) [vertexW] {$(2,2)$};
\node (x23) at (5,3) [vertexY] {$(2,3)$};
\node (x31) at (1,1) [vertexY] {$(3,1)$};
\node (x32) at (3,1) [vertexW] {$(3,2)$};
\node (x33) at (5,1) [vertexY] {$(3,3)$};

\draw [->,line width=0.06cm] (x31) -- (x21);
\draw [->,line width=0.06cm] (x23) -- (x33);

\draw [->,line width=0.06cm] (x21) to [out=30,in=150] (x23);
\draw [->,line width=0.06cm] (x33) to [out=210,in=330] (x31);
\end{tikzpicture} \hfill
%

\end{center}
\caption{Illustrating a cycle in the normal form of $P_{1}$ from Example~\ref{exInitial}. 
}
\label{fig:DP1essential}
\end{figure}
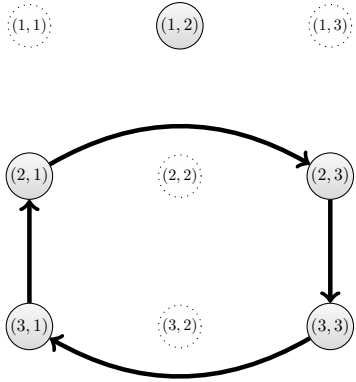

In the next section we formally introduce the directed graph formulation of matching problems due to \cite{Maffray:1992:JCTB}. While the reader not interested in the proofs of our results can skip this section, we believe that reformulating stable matching problems in this way is useful as it allows one to visualise the problem in question.
To illustrate how visually intuitive this equivalent formulation is, Figure~\ref{fig:DP1essential} depicts the normal form of instance $P_{1}$, $P_{1}''$, from Example \ref{exInitial}. With three workers and three firms there is a $3\times3$ grid, where each vertex in the grid is a pair with vertex $(i, j)$ corresponding to the pair $(\worker{i}, \firm{j})$.
A matching is a subset of three vertices no two of which are in the same row nor the same column.

The vertices that are hollow represent pairs that cannot be part of any stable matching as both participants in that pair deemed the other unattractive at some point along the IDUA procedure.
The directed edges (arcs) between pairs that remain depict preferences. For example, the arc from vertex $(2, 1)$ to $(2, 3)$ indicates that worker $\worker{2}$ prefers firm $\firm{3}$ to firm $\firm{1}$ (i.e., $\firm{3} \pref{\worker{2}} \firm{1}$). The clockwise cycle of arcs is easily identified by simple eyeballing, and the reader can verify that this cycle corresponds precisely to that given in \eqref{eq:cycle}. (In the directed graph representing the original instance $P_{1}$, there were many more directed edges, but when a vertex in the directed graph is deleted so too are all directed edges that are incident to it.)

\section{Stable matchings and directed graphs}\label{sec:Digraphs}

In Section~\ref{sec:terminology} we introduce terminology for digraphs.
The terminology is standard and so this section can be skipped by any reader familiar with digraphs.
Section~\ref{sec:matchingDigraphs} introduces the equivalent matching digraph representation of a matching problem introduced by \cite{Maffray:1992:JCTB} and illustrates how the IDUA procedure works via an example.

\subsection{Digraph terminology and notation}\label{sec:terminology}
A {\em directed graph} (or just {\em digraph}) $D$ consists of a non-empty finite set $V (D)$ of elements called {\em vertices} and a finite set $A(D)$ of ordered pairs of distinct vertices called {\em arcs}. We shall call $V(D)$ the {\em vertex set} and $A(D)$ the {\em arc set} of $D$ and write $D=(V(D),A(D)).$ For an arc $xy$ the first vertex $x$ is its {\em tail} and the second vertex $y$ is its {\em head}. Moreover, $x$ is called an {\em in-neighbour} of $y$ and $y$ an {\em out-neighbour} of $x.$ We also say that the arc $xy$ {\em leaves} $x$ and {\em enters} $y$. 
We say that a vertex $x$ is {\em incident} to an arc $a$ if $x$ is the head or tail of $a$. For a vertex $v \in V(D)$, the {\em out-degree} of $v$ in $D$, $d^{+}_{D}(v)$, is the number of out-neighbours of $v$
Similarly, the {\em in-degree} of $v$ in $D$, $d^{-}_{D}(v)$, is the number of in-neighbours of $v$.
A vertex $u$ is {\em isolated} if $d^{+}_{D}(u)=d^{-}_{D}(u)=0.$

A {\it walk}, $W$, in a digraph $D$ is a sequence of vertices $x_1,x_2,\dots , x_p$ for which there is an arc from each vertex in the sequence to its successor.
A walk is written as $W = x_1x_2\dots x_p$.
Special cases of walks are paths and cycles.
A walk $W$ is a {\em path} if the vertices of $W$ are distinct.
If the vertices $x_1, x_2, \dots, x_{p-1}$ are distinct, for $p \geq 2$ and $x_1 = x_p$, then $W$ is a {\em cycle}.

For a digraph $D=(V,A)$ and an arc $xy\in A$, {\em deletion of $xy$ from} $D$ results in the digraph $D-xy=(V,A\setminus \{xy\}).$ For a vertex $v\in V$, {\em deletion of $v$ from} $D$ results in the digraph $D-v=(V\setminus \{v\},A\setminus A_v),$ where $A_v$ is the set of arcs in $A$ incident to $v$. A digraph $D'$ is called a {\em subdigraph} of $D$ if $D'$ is obtained from $D$ by deleting some vertices and arcs. 
If only vertices are deleted, $D'$ is an {\em induced subdigraph} of $D.$

For a textbook treatment of digraphs, see \cite{Bang-JensenGutin:2009:}.

\subsection{Matching digraphs}\label{sec:matchingDigraphs}

Given an instance of the matching problem, $P$, we define the associated {\em matching digraph}, $D(P) = (V, A)$, where $V$ is the vertex set and $A$ is the arc set. The vertex set $V$ is defined as $V := \Workers \times \Firms$. The arc set $A$ is defined as follows.

\begin{center}
$\begin{array}{rcl}
  A_{\Workers} & := & \{(\worker{}, \firm{i}) (\worker{}, \firm{j}) \; | \; \firm{j} \pref{\worker{}} \firm{i} \} \\
  A_{\Firms}  & := &  \{ (\worker{i}, \firm{}) (\worker{j}, \firm{}) \; | \; \worker{j} \pref{\firm{}} \worker{i} \} \\
  A & := & A_{\Workers} \cup A_{\Firms} \\
\end{array}$
\end{center}

{A matching in $P$ is depicted in $D(P)$ by a set of vertices, $M$, such that for every $\worker{} \in \Workers{}$ there exists exactly one vertex, $(\worker{},\firm{}')$, in $M$ containing $\worker{}$ and for every $\firm{} \in \Firms{}$ there exists exactly one vertex, $(\worker{}',\firm{})$, in $M$ containing $\firm{}$. (Going forward we will abuse terminology and refer to such a collection of vertices in $D(P)$ as a matching.)}

A stable matching in $D(P)$ is a matching $M$ such that for every vertex $(\worker{},\firm{}) \in V\big(D(P)\big)$ either $(\worker{},\firm{}) \in M$ or $(\worker{},\firm{})$ has an out-neighbour that belongs to $M$. In the language of directed graphs, $M$ is a {\em kernel}.\footnote{Kernels were first introduced in \cite{NeumannMorgenstern:1944:} as the generalisation of solutions to cooperative games.}

The following example serves three purposes. First, we use it to introduce matching digraphs in a more rigorous manner than at the end of Section \ref{sec:Results}.
Second, the preference lists possess a cycle (in fact more than one), so the acyclicity condition that is sufficient for uniqueness cannot be invoked to conclude that there is a unique stable matching.
And yet there is a unique stable matching. Third, given that there is a unique stable matching, part \ref{thmStatementSingleton} of Theorem \ref{thm:uniqueStableE} confirms that the IDUA will collapse each market participant's preference list to a singleton.
We will use this fact to illustrate how IDUA operates, and to show how visually intuitive the procedure is when a matching problem is reformulated using digraphs.

\begin{example} \label{exUnique}
Let $P_{2}$ be an instance of the matching problem in which there are three workers and three firms. That is, $\Workers = \set{\worker{1}, \worker{2}, \worker{3}}$ and $\Firms = \set{\firm{1}, \firm{2}, \firm{3}}$. Preferences are as follows:
\begin{align*}
\worker{1}: \firm{3} \pref{\worker{1}} \firm{1} \pref{\worker{1}} \firm{2} & \hspace{.4in}  \firm{1}:   \worker{1} \pref{\firm{1}} \worker{2} \pref{\firm{1}} \worker{3}   \\
\worker{2}: \firm{1} \pref{\worker{2}} \firm{2} \pref{\worker{2}} \firm{3}  & \hspace{.4in}  \firm{2}:   \worker{3} \pref{\firm{2}} \worker{1} \pref{\firm{2}} \worker{2}                    \\
\worker{3}: \firm{1} \pref{\worker{3}} \firm{3} \pref{\worker{3}} \firm{2}  & \hspace{.4in}  \firm{3}:   \worker{3} \pref{\firm{3}} \worker{2} \pref{\firm{3}} \worker{1}
\end{align*}

\end{example}

It can be checked that $P_{2}$ possesses a unique stable matching, $\match^{*}$.
It is
\begin{equation}\label{eq:example2stable}
\match^{*}:  \hspace{.3in} (\worker{1}, \firm{1}), (\worker{2}, \firm{2}), (\worker{3}, \firm{3})
\end{equation}
We emphasise that preferences of $P_{2}$ are not acyclic. To see this, consider the subpopulation $\set{\worker{1}, \worker{2}, \firm{1}, \firm{3}}$. The cycle here is given by,
\begin{equation}\label{eq:cycleUnique}
\firm{3} \pref{\worker{1}} \firm{1} \text{ and } \worker{2} \pref{\firm{3}} \worker{1} \text{ and } \firm{1} \pref{\worker{2}} \firm{3} \text{ and } \worker{1} \pref{\firm{1}} \worker{2}
\end{equation}
A relabelling of participants in \eqref{eq:cycleUnique}, as was done in going from \eqref{eq:cycle} to \eqref{eq:cycleOrdered}, confirms the cycle.

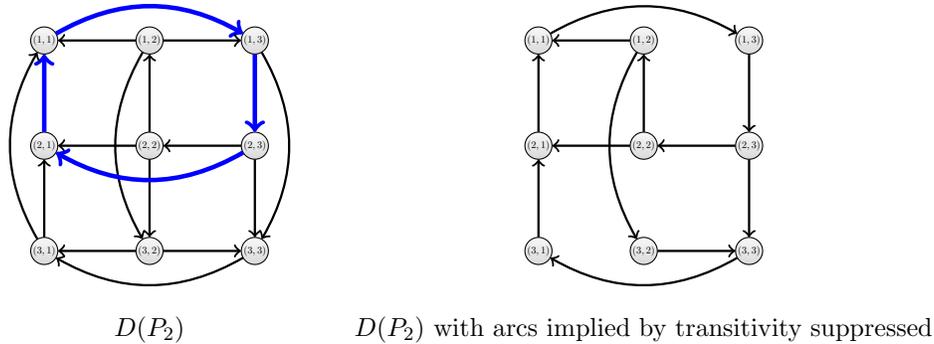
\begin{figure}[hbtp]
\begin{center}
\tikzstyle{vertexDOT}=[scale=0.25,circle,draw,fill]
\tikzstyle{vertexY}=[circle,draw, top color=gray!5, bottom color=gray!30, minimum size=11pt, scale=0.35, inner sep=0.99pt]
\begin{tikzpicture}[scale=0.7]
\node (x11) at (1,5) [vertexY] {$(1,1)$};
\node (x12) at (3,5) [vertexY] {$(1,2)$};
\node (x13) at (5,5) [vertexY] {$(1,3)$};
\node (x21) at (1,3) [vertexY] {$(2,1)$};
\node (x22) at (3,3) [vertexY] {$(2,2)$};
\node (x23) at (5,3) [vertexY] {$(2,3)$};
\node (x31) at (1,1) [vertexY] {$(3,1)$};
\node (x32) at (3,1) [vertexY] {$(3,2)$};
\node (x33) at (5,1) [vertexY] {$(3,3)$};

\draw [->,line width=0.03cm] (x31) -- (x21);
\draw [->,line width=0.06cm, blue] (x21) -- (x11);
\draw [->,line width=0.03cm] (x22) -- (x12);
\draw [->,line width=0.03cm] (x12) to [out=240,in=120] (x32);
\draw [->,line width=0.06cm, blue] (x13) -- (x23);
\draw [->,line width=0.03cm] (x23) -- (x33);

\draw [->,line width=0.03cm] (x12) -- (x11);
\draw [->,line width=0.06cm, blue] (x11) to [out=30,in=150] (x13);
\draw [->,line width=0.03cm] (x23) -- (x22);
\draw [->,line width=0.03cm] (x22) -- (x21);
\draw [->,line width=0.03cm] (x32) -- (x33);
\draw [->,line width=0.03cm] (x33) to [out=210,in=330] (x31);
\draw [->,line width=0.03cm] (x31) to [out=120,in=240] (x11);
\draw [->,line width=0.03cm] (x22) -- (x32);
\draw [->,line width=0.03cm] (x13) to [out=300,in=60] (x33);

\draw [->,line width=0.03cm] (x12) -- (x13);
\draw [->,line width=0.06cm, blue] (x23) to [out=210,in=330] (x21);
\draw [->,line width=0.03cm] (x32) -- (x31);
\node [scale=0.9] at (3,-0.5) {$D(P_{2})$};
\end{tikzpicture} \hfill
\begin{tikzpicture}[scale=0.7]
\node (x11) at (1,5) [vertexY] {$(1,1)$};
\node (x12) at (3,5) [vertexY] {$(1,2)$};
\node (x13) at (5,5) [vertexY] {$(1,3)$};
\node (x21) at (1,3) [vertexY] {$(2,1)$};
\node (x22) at (3,3) [vertexY] {$(2,2)$};
\node (x23) at (5,3) [vertexY] {$(2,3)$};
\node (x31) at (1,1) [vertexY] {$(3,1)$};
\node (x32) at (3,1) [vertexY] {$(3,2)$};
\node (x33) at (5,1) [vertexY] {$(3,3)$};

\draw [->,line width=0.03cm] (x31) -- (x21);
\draw [->,line width=0.03cm] (x21) -- (x11);
\draw [->,line width=0.03cm] (x22) -- (x12);
\draw [->,line width=0.03cm] (x12) to [out=240,in=120] (x32);
\draw [->,line width=0.03cm] (x13) -- (x23);
\draw [->,line width=0.03cm] (x23) -- (x33);

\draw [->,line width=0.03cm] (x12) -- (x11);
\draw [->,line width=0.03cm] (x11) to [out=30,in=150] (x13);
\draw [->,line width=0.03cm] (x23) -- (x22);
\draw [->,line width=0.03cm] (x22) -- (x21);
\draw [->,line width=0.03cm] (x32) -- (x33);
\draw [->,line width=0.03cm] (x33) to [out=210,in=330] (x31);
\node [scale=0.9] at (3,-0.5) {$D(P_{2})$ with arcs implied by transitivity suppressed};
\end{tikzpicture} \hfill
%

\end{center}
\caption{The matching digraph $D(P_{2})$ for the instance $P_{2}$ of Example~\ref{exUnique}. 
}
\label{fig:DP}
\end{figure}

Figure~\ref{fig:DP} illustrates the matching digraph for $P_{2}$. The complete digraph $D(P_{2})$ is displayed in the left hand panel, while the digraph in the right hand panel is a ``condensed'' version where the arcs implied by transitivity have been suppressed for readability.

The vertex $(i,j)$ in each digraph of Figure \ref{fig:DP} denotes the pair $(\worker{i}, \firm{j})$. That is, rows are indexed by workers and columns are indexed by firms. The preferences of workers are depicted by horizontal arcs, the arc set $A_{\Workers}$, and the preferences of firms are depicted by the vertical arcs, the arc set $A_{\Firms}$. A matching in $P_{2}$ corresponds to a set of vertices, $M$, no two of which are in the same row nor the same column.

Since every participant's preference is transitive, there cannot be a cycle in any row (corresponding to a worker's preferences) nor in any column (corresponding to a firm's preferences) of a matching digraph. Note however that there can be cycles in the digraph as a whole. In Figure \ref{fig:DP}, we have labelled one such preference cycle by colouring the arcs that comprise it in \blue{blue}, and we note that this is precisely the preference cycle identified in \eqref{eq:cycleUnique}. So the acyclic condition of \cite{Chung:2000:GEB}, specified in Definition \ref{def:cycle}, requires that preferences are intertwined in such a way that the transitivity property of individual preferences is inherited at the population level.\footnote{In fact, Example~\ref{exUnique} is a counterexample to Lemma 6 and Corollary 7 in \cite{Eeckhout:2000:EL}. That is, the $n=3$ matching problem in Example~\ref{exUnique} possesses a unique stable matching and yet:\ (i) as seen in Figure~\ref{fig:DP}, there is a preference cycle (indicating that Eeckhout's Lemma 6 is not correct as stated), and (ii) no worker is matched to their most preferred firm (violating Eeckhout's Corollary 7).}

Given a stable matching problem $P$ and its associated matching digraph, $D(P)$, we now introduce a reduction, $\reduce$, that ``prunes'' the matching digraph of extraneous information.
Specifically, it identifies and deletes vertices in $D(P)$ that represent worker-firm pairs that view each other as unattractive.

For every vertex $v \in V\big(D(P)\big)$, it will be useful to decompose $d^{+}_{D}(v)$ into $d^{+}_{D}(v) = d^{+}_{\Workers}(v) + d^{+}_{\Firms}(v)$. That is, the out-degree of a vertex is split into the horizontal out-degree and the vertical out-degree. (Note that for for a given pair $(\worker{}, \firm{})$, the horizontal out-degree corresponds to the number of firms that $\worker{}$ prefers to $\firm{}$ and the vertical out-degree corresponds to the number of workers that $\firm{}$ prefers to $\worker{}$.)
We now have the following.

\begin{definition}\label{def:reduce}
Given a matching digraph $D(P)$, we define $\reduce\big(D(P)\big)$ as the result of the following procedure.

Choose $v = (\worker{}, \firm{}) \in V(D(P))$  with either $d^{+}_{\Firms}(v) = 0$ or $d^{+}_{\Workers}(v) = 0$.
If $d^{+}_{\Firms}(v) = 0$, delete all vertices $(\worker{}, \firm{i})$ such that $(\worker{}, \firm{i}) (\worker{}, \firm{})  \in A_{\Workers}$.
Otherwise (i.e., $d^{+}_{\Workers}(v) = 0$), delete all vertices $(\worker{i}, \firm{})$ such that $(\worker{i}, \firm{}) (\worker{}, \firm{})  \in A_{\Firms}$.\footnote{{If more than one vertex $(\worker{}, \firm{}) \in V\big(D(P)\big)$ satisfies the condition above, then $\reduce(D(P))$ will depend on which of them is chosen. However, we can ignore this fact since our interest lies in repeated application of $\reduce$ (see below) and, as per Lemma \ref{lemma:orderIndependentE}, the end result of repeated application does not depend on the intermediate values.}}
\end{definition}

The reduction procedure, $\reduce$, operates as follows. If $d^{+}_{\Firms}(v) = 0$, then all vertices that are the tail of an arc with head $v$ in $A_{\Workers}$ are to be deleted. This is because if $v = (\worker{}, \firm{})$, and $d^{+}_{\Firms}(v) = 0$, then worker $\worker{}$ is firm $\firm{}$'s most preferred worker and hence $\worker{}$ can not be matched with any firm that they prefer less than $\firm{}$ in any stable matching. When $d^{+}_{\Workers}(v) = 0$, analogous vertex deletions are performed. 

A version of IDUA, which we call IDUA$^{\reduce}$, repeats $\reduce$ after setting $D(P):=\reduce\big(D(P)\big)$ until further reductions are no longer possible. When IDUA$^{\reduce}$ stops we obtain the normal form of the initial $D(P)$ denoted by  $D^{*}(P).$ By definition,  $D^{*}(P) = D(P^{*})$.

We emphasise that the reduction procedure IDUA$^{\reduce}$, generated by repeated applications of $\reduce$, differs slightly from the IDUA procedure of Definition~\ref{def:IDUAe}.
The difference is as follows.
IDUA, as defined in Definition~\ref{def:IDUAe}, aligns more closely with IDDS in strategic games, as it involves multiple simultaneous deletions.\footnote{While algorithmically more care must be taken with operations that delete multiple objects simultaneously, simultaneous deletion procedures are a closer fit to the higher ordering reasoning systems that game theorists assume of rational agents.}
Specifically, in iteration $k$, IDUA deletes all vertices that can be identified as not part of some stable matching. It does this by pivoting around all blocking pairs simultaneously. This should be contrasted with the IDUA$^{\reduce}$, that in iteration $k$ finds one blocking pair and pivots only around it. But while there is a formal difference between IDUA and IDUA$^{\reduce}$, from a practical perspective the difference is immaterial as Lemma \ref{lemma:orderIndependentE} confirms that the order in which unattractive alternatives are deleted does not affect the final output.

\begin{figure}[hbtp]
\begin{center}
\begin{tabular}{|c|c|c|} \hline
\tikzstyle{vertexDOT}=[scale=0.25,circle,draw,fill]
\tikzstyle{vertexY}=[circle,draw, top color=gray!5, bottom color=gray!30, minimum size=11pt, scale=0.3, inner sep=0.99pt]
\tikzstyle{vertexZ}=[circle,draw, top color=black!40, bottom color=black!80, minimum size=11pt, scale=0.3, inner sep=0.99pt]
\tikzstyle{vertexW}=[circle,dotted, draw, top color=gray!1, bottom color=gray!1, minimum size=11pt, scale=0.3, inner sep=0.99pt]
\begin{tikzpicture}[scale=0.6]
\node [scale=0.9] at (6.3,3) {$\red{\leftarrow}$};

\node (x11) at (1,5) [vertexY] {$(1,1)$};
\node (x12) at (3,5) [vertexY] {$(1,2)$};
\node (x13) at (5,5) [vertexY] {$(1,3)$};
\node (x21) at (1,3) [vertexZ] {$(2,1)$};
\node (x22) at (3,3) [vertexY] {$(2,2)$};
\node (x23) at (5,3) [vertexY] {$(2,3)$};
\node (x31) at (1,1) [vertexW] {$(3,1)$};
\node (x32) at (3,1) [vertexY] {$(3,2)$};
\node (x33) at (5,1) [vertexY] {$(3,3)$};

\draw [->,dotted, line width=0.03cm] (x31) -- (x21);
\draw [->,line width=0.03cm] (x21) -- (x11);
\draw [->,line width=0.03cm] (x22) -- (x12);
\draw [->,line width=0.03cm] (x12) to [out=240,in=120] (x32);
\draw [->,line width=0.03cm] (x13) -- (x23);
\draw [->,line width=0.03cm] (x23) -- (x33);

\draw [->,line width=0.03cm] (x12) -- (x11);
\draw [->,line width=0.03cm] (x11) to [out=30,in=150] (x13);
\draw [->,line width=0.03cm] (x23) -- (x22);
\draw [->,line width=0.03cm] (x22) -- (x21);
\draw [->,line width=0.03cm] (x32) -- (x33);
\draw [->,dotted,line width=0.03cm] (x33) to [out=210,in=330] (x31);
\draw [->,dotted,line width=0.03cm] (x31) to [out=120,in=240] (x11);
\draw [->,line width=0.03cm] (x22) -- (x32);
\draw [->,line width=0.03cm] (x13) to [out=300,in=60] (x33);

\draw [->,line width=0.03cm] (x12) -- (x13);
\draw [->,line width=0.03cm] (x23) to [out=210,in=330] (x21);
\draw [->,dotted,line width=0.03cm] (x32) -- (x31);
\end{tikzpicture} & 
\tikzstyle{vertexDOT}=[scale=0.25,circle,draw,fill]
\tikzstyle{vertexY}=[circle,draw, top color=gray!5, bottom color=gray!30, minimum size=11pt, scale=0.3, inner sep=0.99pt]
\tikzstyle{vertexZ}=[circle,draw, top color=black!40, bottom color=black!80, minimum size=11pt, scale=0.3, inner sep=0.99pt]
\tikzstyle{vertexW}=[circle,dotted, draw, top color=gray!1, bottom color=gray!1, minimum size=11pt, scale=0.3, inner sep=0.99pt]
\begin{tikzpicture}[scale=0.6]
\node [scale=0.9] at (6.1,1) {$\red{\leftarrow}$};

\node (x11) at (1,5) [vertexY] {$(1,1)$};
\node (x12) at (3,5) [vertexY] {$(1,2)$};
\node (x13) at (5,5) [vertexW] {$(1,3)$};
\node (x21) at (1,3) [vertexY] {$(2,1)$};
\node (x22) at (3,3) [vertexY] {$(2,2)$};
\node (x23) at (5,3) [vertexW] {$(2,3)$};
\node (x32) at (3,1) [vertexY] {$(3,2)$};
\node (x33) at (5,1) [vertexZ] {$(3,3)$};

\draw [->,line width=0.03cm] (x21) -- (x11);
\draw [->,line width=0.03cm] (x22) -- (x12);
\draw [->,line width=0.03cm] (x12) to [out=240,in=120] (x32);
\draw [->,dotted,line width=0.03cm] (x13) -- (x23);
\draw [->,dotted,line width=0.03cm] (x23) -- (x33);

\draw [->,line width=0.03cm] (x12) -- (x11);
\draw [->,dotted,line width=0.03cm] (x11) to [out=30,in=150] (x13);
\draw [->,dotted,line width=0.03cm] (x23) -- (x22);
\draw [->,line width=0.03cm] (x22) -- (x21);
\draw [->,line width=0.03cm] (x32) -- (x33);
\draw [->,line width=0.03cm] (x22) -- (x32);
\draw [->,dotted,line width=0.03cm] (x13) to [out=300,in=60] (x33);

\draw [->,dotted,line width=0.03cm] (x12) -- (x13);
\draw [->,dotted,line width=0.03cm] (x23) to [out=210,in=330] (x21);
\end{tikzpicture} & 
\tikzstyle{vertexDOT}=[scale=0.25,circle,draw,fill]
\tikzstyle{vertexY}=[circle,draw, top color=gray!5, bottom color=gray!30, minimum size=11pt, scale=0.3, inner sep=0.99pt]
\tikzstyle{vertexZ}=[circle,draw, top color=black!40, bottom color=black!80, minimum size=11pt, scale=0.3, inner sep=0.99pt]
\tikzstyle{vertexW}=[circle,dotted, draw, top color=gray!1, bottom color=gray!1, minimum size=11pt, scale=0.3, inner sep=0.99pt]
\begin{tikzpicture}[scale=0.6]
\node [scale=0.9] at (4.0,5) {$\red{\leftarrow}$};

\node (x11) at (1,5) [vertexZ] {$(1,1)$};
\node (x12) at (3,5) [vertexY] {$(1,2)$};
\node (x21) at (1,3) [vertexW] {$(2,1)$};
\node (x22) at (3,3) [vertexY] {$(2,2)$};
\node (x32) at (3,1) [vertexY] {$(3,2)$};
\node (x33) at (5,1) [vertexY] {$(3,3)$};

\draw [->,dotted,line width=0.03cm] (x21) -- (x11);
\draw [->,line width=0.03cm] (x22) -- (x12);
\draw [->,line width=0.03cm] (x12) to [out=240,in=120] (x32);
\draw [->,line width=0.03cm] (x12) -- (x11);
\draw [->,dotted,line width=0.03cm] (x22) -- (x21);
\draw [->,line width=0.03cm] (x32) -- (x33);
\draw [->,line width=0.03cm] (x22) -- (x32);
\end{tikzpicture} \\ \hline
\tikzstyle{vertexDOT}=[scale=0.25,circle,draw,fill]
\tikzstyle{vertexY}=[circle,draw, top color=gray!5, bottom color=gray!30, minimum size=11pt, scale=0.3, inner sep=0.99pt]
\tikzstyle{vertexZ}=[circle,draw, top color=black!40, bottom color=black!80, minimum size=11pt, scale=0.3, inner sep=0.99pt]
\tikzstyle{vertexW}=[circle,dotted, draw, top color=gray!1, bottom color=gray!1, minimum size=11pt, scale=0.3, inner sep=0.99pt]
\begin{tikzpicture}[scale=0.6]
\node [scale=0.9] at (3,5.6) {\mbox{ }};
\node [scale=0.9] at (3,0.4) {\mbox{ }};
\node [scale=0.9] at (1,2) {$\red{\uparrow}$};

\node (x11) at (1,5) [vertexZ] {$(1,1)$};
\node (x12) at (3,5) [vertexW] {$(1,2)$};
\node (x22) at (3,3) [vertexY] {$(2,2)$};
\node (x32) at (3,1) [vertexY] {$(3,2)$};
\node (x33) at (5,1) [vertexY] {$(3,3)$};

\draw [->,dotted,line width=0.03cm] (x22) -- (x12);
\draw [->,dotted,line width=0.03cm] (x12) to [out=240,in=120] (x32);
\draw [->,dotted,line width=0.03cm] (x12) -- (x11);
\draw [->,line width=0.03cm] (x32) -- (x33);
\draw [->,line width=0.03cm] (x22) -- (x32);
\end{tikzpicture} & 
\tikzstyle{vertexDOT}=[scale=0.25,circle,draw,fill]
\tikzstyle{vertexY}=[circle,draw, top color=gray!5, bottom color=gray!30, minimum size=11pt, scale=0.3, inner sep=0.99pt]
\tikzstyle{vertexZ}=[circle,draw, top color=black!40, bottom color=black!80, minimum size=11pt, scale=0.3, inner sep=0.99pt]
\tikzstyle{vertexW}=[circle,dotted, draw, top color=gray!1, bottom color=gray!1, minimum size=11pt, scale=0.3, inner sep=0.99pt]
\begin{tikzpicture}[scale=0.6]
\node [scale=0.9] at (3,5.6) {\mbox{ }};
\node [scale=0.9] at (3,0.4) {\mbox{ }};
\node [scale=0.9] at (5,4) {$\red{\downarrow}$};

\node (x11) at (1,5) [vertexY] {$(1,1)$};
\node (x22) at (3,3) [vertexY] {$(2,2)$};
\node (x32) at (3,1) [vertexW] {$(3,2)$};
\node (x33) at (5,1) [vertexZ] {$(3,3)$};

\draw [->,dotted,line width=0.03cm] (x32) -- (x33);
\draw [->,dotted,line width=0.03cm] (x22) -- (x32);
\end{tikzpicture} & 
\tikzstyle{vertexDOT}=[scale=0.25,circle,draw,fill]
\tikzstyle{vertexY}=[circle,draw, top color=gray!5, bottom color=gray!30, minimum size=11pt, scale=0.3, inner sep=0.99pt]
\tikzstyle{vertexZ}=[circle,draw, top color=black!40, bottom color=black!80, minimum size=11pt, scale=0.3, inner sep=0.99pt]
\tikzstyle{vertexW}=[circle,dotted, draw, top color=gray!1, bottom color=gray!1, minimum size=11pt, scale=0.3, inner sep=0.99pt]
\begin{tikzpicture}[scale=0.6]
\node [scale=0.9] at (3,5.6) {\mbox{ }};
\node [scale=0.9] at (3,0.4) {\mbox{ }};

\node (x11) at (1,5) [vertexY] {$(1,1)$};
\node (x22) at (3,3) [vertexY] {$(2,2)$};
\node (x33) at (5,1) [vertexY] {$(3,3)$};
\end{tikzpicture} \\ \hline
\end{tabular}
\end{center}
\caption{An illustration of how IDUA$^{\reduce}$ operates.}
\label{fig:reductions}
\end{figure}
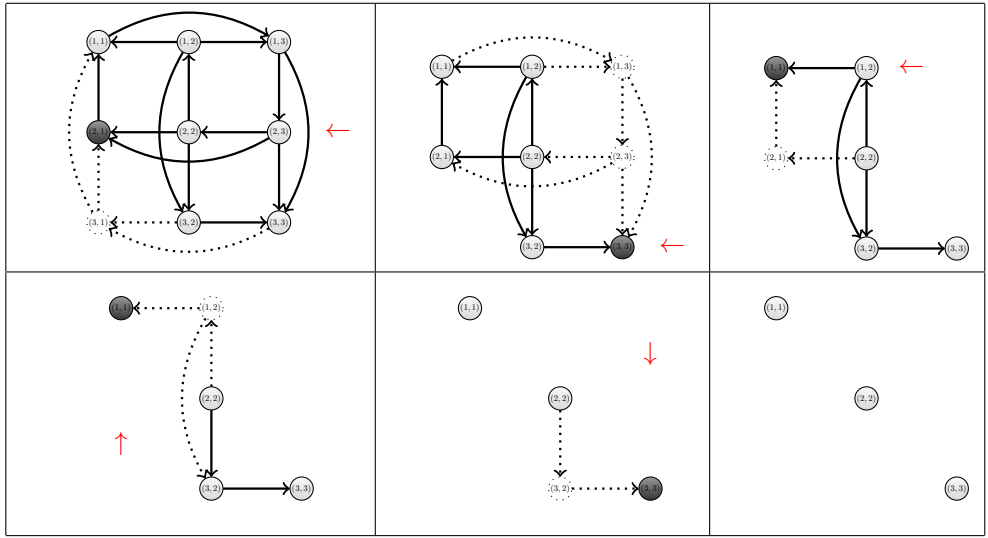

We now illustrate how IDUA$^{\reduce}$ operates, using instance $P_{2}$ from Example~\ref{exUnique}. (Let us recall that these preferences are not acyclic and yet this instance does possess a unique stable matching.) Figure~\ref{fig:reductions} below contains six panels that show repeated applications of $\reduce$ to the matching digraph of $P_{2}$, $D(P_{2})$. As in Figure~\ref{fig:DP}, vertex $(i,j)$ denotes the pair $(\worker{i}, \firm{j})$. In each panel the red arrow indicates the row or column where $\reduce$ is being applied.
The black vertex is the vertex with no arcs out of it in the row or column that is being considered (i.e., it is the blocking pair that $\reduce$ pivots around). The dotted vertices and arcs are the vertices and arcs that get deleted in that iteration (the dotted vertices have arcs into the black vertex, and these arcs are perpendicular to the row/column that determined the choice of the black vertex). Once the repeated application can go no further, we have the normal form of $D(P_{2})$, $D^*(P_{2})$, that consists of the matching in \eqref{eq:example2stable}.

Let us now be slightly more concrete about Figure~\ref{fig:reductions}. In the first panel, we consider the second row that captures worker $\worker{2}$'s preferences. The (horizontal) arcs in this row indicate that $\firm{1}$ is $\worker{2}$'s most preferred firm, and so we pivot around vertex $(\worker{2}, \firm{1})$. We indicate this by colouring this vertex black. Firm $\firm{1}$ recognises that worker $\worker{2}$ is, in effect, a lower bound on who they can pair with. (Since if firm $\firm{1}$ were matched with any worker that they prefer less than worker $\worker{2}$, then $(\worker{2}, \firm{1})$ would constitute a blocking pair since $\worker{2}$ will leave any firm for $\firm{1}$.) Worker $\worker{3}$ is one such worker. So firm $\firm{1}$ deems $\worker{3}$ unattractive and by the reciprocal nature of this relation, $\worker{3}$ deems $\firm{1}$ unattractive. Hence by considering the second row, we can delete the vertex $(\worker{3}, \firm{1})$ and any arcs incident on it.

In the second panel, consider the third row representing worker $\worker{3}$'s preferences.
At the onset, $\firm{1}$ was worker $\worker{3}$'s most preferred firm. But the first application of $\reduce$, depicted in the first panel, showed that the pair $(\worker{3}, \firm{1})$ can never be part of a stable matching as vertex $(\worker{3}, \firm{1})$ was deleted. As such, $\worker{3}$'s most preferred feasible firm is $\firm{3}$, which we depict by colouring the vertex $(\worker{3}, \firm{3})$ black. This means that firm $\firm{3}$ can at worst match with worker $\worker{3}$. Thus we can delete all vertices with arcs in $A_{F}$ that have head $(\worker{3}, \firm{3})$. (In fact, given that $\worker{3}$ is in fact $\firm{3}$'s most preferred feasible partner, we can now conclude that they will certainly be paired in all stable matchings.) Note that this deletes vertices  $(\worker{1}, \firm{3})$ and $(\worker{2}, \firm{3})$, and doing so this deletes three of the four arcs comprised the cycle in the preference lists, i.e., the arcs that were coloured \blue{blue} in the left hand panel of Figure \ref{fig:DP}. This highlights that cycles are not always a barrier to uniqueness.

The remaining panels should now be easily understood.



\section{Conclusion and extensions for future work}\label{sec:Conclusion}

\paragraph{Conclusion.}\,

\noindent
{The two-sided matching framework of \cite{GaleShapley:1962:AMM} is one of the classic models of economic theory. It has been applied widely to settings ranging from school admissions, entry-level labour markets, refugee resettlement, and others (see \cite{Roth:2008:IJGT} for a survey).
Yet despite all the attention that the model has received, a classification of what structural properties guarantee a unique stable matching remained an open question.
In this paper we have resolved this open question.}

It turns out that the key to answering the above question is first to reduce a matching problem to its essential constituents.
We do this by applying a reduction procedure, the {\em iterated deletion of unattractive alternatives} (IDUA), that strips away the parts of the preference lists that are not relevant to the set of stable matchings. We term the resultant (sub)matching problem the {\it normal form}.
Our main result, Theorem \ref{thm:uniqueStableE}, shows that a matching problem has a unique stable matching if and only if preferences on the normal form are acyclic if and only if the normal form is precisely the unique stable matching and nothing more (i.e., IDUA collapses every market participant's preference list to a singleton).

\paragraph{Extensions and open problems.}\,

\noindent
The matching environment that we have studied in this paper (Definition \ref{def:instance}) is precisely that originally considered by \cite{GaleShapley:1962:AMM}.
Such an environment is defined by the following three features.
\begin{enumerate}[label=(\roman*)]
\item\label{feature:one-to-one}
{\it a one-to-one market}: each worker can take only one job and each firm has only one position.

\item\label{feature:complete}
{\it all preferences are complete}: every worker prefers to be employed at any firm over being unemployed, and, every firm prefers to fill its position with any worker over leaving the position unfilled.

\item\label{feature:balanced}
{\it a balanced market}: there are exactly the same number of workers as firms/positions.

\end{enumerate}

The above assumptions are not universal to all matching markets and so there are many ways in which this original \cite{GaleShapley:1962:AMM} environment has been extended.
Each more general variant involves relaxing some combination of the three features above.
As mentioned previously, our main result, Theorem \ref{thm:uniqueStableE}, holds if preferences are incomplete (relaxing feature \ref{feature:complete} above) and the market is unbalanced (relaxing feature \ref{feature:balanced} above).
The caveat is that there may be some market participants that are not included in the stable matching.
(Furthermore, the definition of the environment, Definition~\ref{def:instance}, of a blocking pair, Definition~\ref{def:blocking}, and of a stable matching, Definition~\ref{def:stableMatch}, all have to be amended. And unfortunately doing so comes at great notational cost.)

In \cite{GutinNeary:2022:arXiv} we consider many-to-one matching markets.
While our focus there is not the issue of uniqueness, it is quite clear that the uniqueness result of this paper extends to that of an unbalanced many-to-one matching market in which preferences need not be complete.\footnote{\cite{GutinNeary:2022:arXiv} is, to the best of our knowledge, the first paper to use the digraph approach for the study of many-to-one matching environments.}
This confirms that partially relaxing feature \ref{feature:one-to-one} is also possible. (Though again there is the caveat that not all workers be employed and/or not all positions be filled.)
The most general environment, the so-called many-to-many matching markets, presents a greater challenge as there is no uniformly agreed upon notion of stability \citep{Sotomayor:1999:MSS, EcheniqueOvideo:2006:TE, KonishiUnver:2006:JET}.
Thus, if and how our uniqueness result carries over to many-to-many matching markets is an open problem.

\cite{AshlagiKanoria:2017:JPE} consider a variant of the matching problem with feature \ref{feature:balanced} above relaxed.
Precisely, they consider an unbalanced one-to-one market with $n$ workers and $n+1$ firms where all preferences are complete. There are precisely $((n+1)!)^{n}(n!)^{n+1}$ possible instances of such an environment.
Amongst other things, \cite{AshlagiKanoria:2017:JPE} show that as $n$ tends to infinity the fraction of instances with a unique stable matching tends to 1.
They attribute this finding to a consequence of increased {\it competition}.

Our classification of unique stable matchings complements this result of \cite{AshlagiKanoria:2017:JPE} in the following way.
If their result is a statement about {\it how} the core becomes small as an unbalanced market size grows, then our Theorem \ref{thm:uniqueStableE} is a statement about {\it why} the core becomes small as an unbalanced market size grows.
That is, as the size of unbalanced markets grows without bound, in most instances the preferences on the normal form become acyclic.
This generates a puzzle.
The normal form of an unbalanced $n \times (n+1)$ matching market with complete preferences is a balanced $n \times n$ market.
So how is it that an additional market participant typically induces an acyclic normal form as $n$ gets large?


\newpage

\appendix

\section*{APPENDIX}\label{APP}


\section{Proofs omitted from the main text}\label{app:proofs}

Lemma~\ref{lemma:iduaSameMatchingsE}, Lemma~\ref{lemma:orderIndependentE}, Lemma~\ref{lemma:extremalMatchesE}, and Theorem \ref{thm:uniqueStableE} are each stated in the text in terms of the standard primitives of the matching model:\ preferences.
Since all of our results are proved using the equivalent matching digraph formulation, below we first restate each result as it appeared in the main text and we then state the equivalent result in terms of matching digraphs. That is, Lemma~\ref{lemma:iduaSameMatchingsE} is reformulated as Lemma~\ref{lemma:iduaSameMatchings}, Lemma~\ref{lemma:orderIndependentE} as Lemma~\ref{lemma:orderIndependent}, Lemma~\ref{lemma:extremalMatchesE} as Lemma~\ref{lemma:extremalMatches}, and Theorem~\ref{thm:uniqueStableE} as Theorem~\ref{thm:uniqueStable}.

\begin{customlemma}{1}
The iterated deletion of unattractive alternatives does not change the set of stable matchings.  That is, $P$ and its associated normal form, $P^{*}$, contain exactly the same set of stable matchings.
\end{customlemma}

\begin{customlemma}{1$'$}\label{lemma:iduaSameMatchings}
The iterated deletion of unattractive alternatives does not change the set of stable matchings.  That is, $D(P)$ and $D^*(P)$ contain exactly the same stable matchings.
\end{customlemma}

\paragraph{Proof of Lemma \ref{lemma:iduaSameMatchings} (and hence of Lemma \ref{lemma:iduaSameMatchingsE}).}
\begin{proof}
We will show that one iteration of $\reduce$ does not change the set of stable matchings.
This will confirm that the lemma holds. 

Assume that  there is no arc in $A_{\Firms}$ leaving $(\worker{},\firm{}) \in V\big(D(P)\big)$  and we have therefore deleted all vertices
$(\worker{}, \firm{i})$ such that $(\worker{}, \firm{i}) (\worker{}, \firm{})  \in A_{\Workers}$ (see (i) in Definition~\ref{def:reduce}).
Let $D_1$ be the matching digraph before the reduction and let $D_2$ denote the matching digraph after the operation.

For the sake of contradiction assume that one of the deleted vertices, say $(\worker{}, \firm{i})$, lies in a stable matching $M_1$ of $D_1$.
Then for some $j,$ we have $(\worker{j}, \firm{})$ is also in $M_1$. Now we will show that $(\worker{}, \firm{})$ is a blocking pair in $M_1.$ Since $(\worker{}, \firm{i})$ and $(\worker{j}, \firm{})$ are in $M_1$,
$(\worker{}, \firm{}) \not\in M_1.$ Moreover, $f$ prefers $w$ to $w_j$ as  $(\worker{j}, \firm{}) (\worker{}, \firm{})  \in A_{\Firms}$ and $w$ prefers $f$ to $f_i$ as $(\worker{}, \firm{i}) (\worker{}, \firm{})  \in A_{\Workers}.$
Therefore, $M_1$ does not exist (as there is a blocking pair).
So no deleted vertex can belong to a stable matching of $D_1$.

Therefore, if $M_1$ is a stable matching in $D_1$, then $M_1$ is also a stable matching in $D_2$ (as $D_2$ is an induced subdigraph of $D_1$ and
$M_1 \subseteq V(D_2)$).  Conversely assume that $M_2$ is a stable matching in $D_2$. Clearly no vertex in $D_2$ is a blocking pair for $M_2$.
For the sake of contradiction assume that $(\worker{}, \firm{i})$ is a blocking pair for $M_2$ in $D_1$ (where $(\worker{}, \firm{i})$ is deleted when 
constructing $D_2$).  Recall that $(\worker{}, \firm{i}) (\worker{}, \firm{})  \in A_{\Workers}$ and there are no arcs in $A_{\Firms}$ leaving
$(\worker{}, \firm{}) \in V(D(P))$.
As $(\worker{}, \firm{})$ is not a blocking pair for $M_2$ in $D_2$, we note that $(\worker{}, \firm{})$ either belongs to $M_2$ or there is a 
vertex $(\worker{}, \firm{j}) \in M_2$, where $(\worker{}, \firm{}) (\worker{}, \firm{j})  \in A_{\Workers}$.
As $(\worker{}, \firm{i}) (\worker{}, \firm{}) \in A_{\Workers}$ and  $(\worker{}, \firm{}) (\worker{}, \firm{j})  \in A_{\Workers}$
we note that $(\worker{}, \firm{i}) (\worker{}, \firm{j})  \in A_{\Workers}$ and therefore  $(\worker{}, \firm{i})$ is not a blocking pair for $M_2$ in $D_1$.
This implies that there is no blocking pair for $M_2$ in $D_1$ and therefore $M_2$ is a stable matching in $D_1$.
This completes the proof.
\end{proof}

\begin{customlemma}{2}
Let $P$ be an instance of the matching problem. Then the normal form of $P$, $P^{*}$, is uniquely defined. That is, no matter in which order we repeatedly delete unattractive alternatives from preference lists, we always end up with the same $P^{*}$.

Furthermore, suppose that $\firm{} \in \set{\pref{\worker{}}^{*}}$ and $\worker{} \in \{\pref{\firm{}}^{*}\}$.
Then, either $(\worker{}, \firm{})$ is part of some stable matching or the following property ($s$) holds.
\begin{description}
\item[($s$):] There exist firms $\firm{j_1}$ and $\firm{j_2}$ such that $\firm{j_1}, \firm{j_2} \in \{\pref{\worker{}}^{*}\}$ with $\firm{j_1} \pref{\worker{}}^{*} \firm{} \pref{\worker{}}^{*} \firm{j_2}$, and there exist workers $\worker{i_1}$ and $\worker{i_2}$ such that $\worker{i_1}, \worker{i_2} \in \{\pref{\firm{}}^{*}\}$ with $\worker{i_1} \pref{\firm{}}^{*} \worker{} \pref{\firm{}}^{*} \worker{i_2}$.
\end{description}
\end{customlemma}

\begin{customlemma}{2$'$}\label{lemma:orderIndependent}
Let $P$ be an instance of the matching problem. Then the matching digraph of the normal form, $D^*(P)$, is uniquely defined. That is, no matter in which order we repeatedly apply $\reduce$ to perform IDUA, we always end up with the same $D^*(P)$.

Furthermore, if $ M \subseteq V\big(D(P)\big)$ denotes the vertices that belong to some stable matching of $P$, 
then $D^*(P)$ contains a vertex $(\worker{},\firm{})$ if and only if
$(\worker{},\firm{}) \in M$ or the following property ($s$) holds.
\begin{description}
\item[($s$):] There exists vertices $(\worker{i_1},\firm{}),(\worker{i_2},\firm{}),(\worker{},\firm{j_1}),(\worker{},\firm{j_2}) \in M$, such that
$(\worker{i_1},\firm{})(\worker{},\firm{})(\worker{i_2},\firm{})$ and $(\worker{},\firm{j_1})(\worker{},\firm{})(\worker{},\firm{j_2})$ are paths in
$V(D(P))$.
\end{description}
\end{customlemma}

\paragraph{Proof of Lemma \ref{lemma:orderIndependent} (and hence of Lemma \ref{lemma:orderIndependentE}).}
\begin{proof}
Let $P$ be an instance that contains a stable matching. 
Let $M \subseteq V(D(P))$ be the vertices of $D(P)$ that belong to some stable matching of $P$.
Let $M^*$ denote the set of vertices $(w,f)\in D(P)$ where either $(w,f)\in M$ or there are vertices $(w,f_i) \in M$ and $(w_j,f) \in M$ such that
$(w,f_i)(w,f)\in A_{\Workers}$ and $(w_j,f)(w,f)\in A_{\Firms}.$
We will show that no matter in what 
order we perform the reductions we always obtain $V\big(D^*(P)\big)=M^*$ and $A\big(D^*(P)\big)$ contain the arcs of $D(P)$ 
with both tail and head in $M^*$.

Let $m^*$ be an arbitrary vertex in $M^*$. If $m^*$ would be deleted by some
Reduction~$\reduce$, then either all vertices incident to arcs in $A_{\Workers}$ that enter $m^*$ will also be deleted
or all vertices incident to arcs in $A_{\Firms}$ that enter $m^*$  will be deleted (by the definition of Reduction~$\reduce$).
In both cases at least one vertex from $M$ will be deleted, a contradiction to Lemma~\ref{lemma:iduaSameMatchings}.
So, if $m^* \in M^*$ then $m^*$ is not deleted by any Reduction~$\reduce$ and $m^* \in V\big(D^*(P)\big)$.
This implies that $M^* \subseteq V\big(D^*(P)\big)$.

For the sake of contradiction assume that there exists a pair $(\worker{},\firm{}) \in V\big(D^*(P)\big) \setminus M^*$.
As $(\worker{},\firm{}) \not\in M^*$ we note that $(\worker{},\firm{}) \not\in M$ and without loss of generality we may assume 
that there is no arc $a\in A_{\Workers{}}$ such that $(\worker{},\firm{})$ is the head of $a$ and the tail of $a$  
 lies in $M$. 
Since $M_{\Firms}$ (see the definition in Lemma \ref{lemma:extremalMatches}) is a stable matching it contains
a vertex $(\worker{},\firm{i}) \in M_{\Firms}$.
As $(\worker{},\firm{})$ has no in-neighbour  of the form $(\worker{},\firm{j})$ belonging to $M$, and therefore also not belonging to $M_{\Firms}$, we have that $(\worker{},\firm{i})$
is an out-neighbour of $(\worker{},\firm{})$. However as there are no arc in $A_{\Firms}$ leaving $(\worker{},\firm{i})$ and
$(\worker{},\firm{})(\worker{},\firm{i})\in A_{\Workers{}},$ we note that $(\worker{},\firm{})$ will be deleted by 
Reduction~$\reduce$, a contradiction to $(\worker{},\firm{}) \in V\big(D^*(P)\big)$. So $V\big(D^*(P)\big)=M^*$ as desired.

We will now prove the second part of the lemma.  If $(\worker{},\firm{}) \in M$ or if $(\worker{},\firm{})$ satisfies Property~($s$) then 
clearly $(\worker{},\firm{}) \in M^* = V\big(D^*(P)\big)$ (due to the existence of $(\worker{i_1},\firm{})$ and $(\worker{},\firm{j_1})$ if Property~($s$) holds).

Conversely let $(\worker{},\firm{}) \in M^*$ be arbitrary. We want to show that $(\worker{},\firm{}) \in M$ or that Property~($s$) holds.
For the sake of contradiction assume that this is not the case. As $(\worker{},\firm{}) \in M^*$ and $(\worker{},\firm{}) \not\in M$ we note that
there exists $(\worker{i_1},\firm{}),(\worker{},\firm{j_1}) \in M$, such that $(\worker{i_1},\firm{})(\worker{},\firm{}) \in A_{\Firms}$ and 
$(\worker{},\firm{j_1})(\worker{},\firm{}) \in A_{\Workers}$. Let $M_{(\worker{i_1},\firm{})}$ denote the stable matching in $P$ containing
$(\worker{i_1},\firm{})$ (which exists as $(\worker{i_1},\firm{}) \in M$). In $M_{(\worker{i_1},\firm{})}$ let $(\worker{},\firm{j_2})$ be the 
vertex containing $\worker{}$. As $(\worker{},\firm{})$ is not a blocking pair in $M_{(\worker{i_1},\firm{})}$ we must have $(\worker{},\firm{})(\worker{},\firm{j_2}) \in A_{\Workers}$ and therefore $(\worker{},\firm{j_1})(\worker{},\firm{})(\worker{},\firm{j_2})$ is a path in $V\big(D(P)\big)$.

Analogously, let  $M_{(\worker{},\firm{j_1})}$ denote the stable matching in $P$ containing
$(\worker{},\firm{j_1})$. In $M_{(\worker{},\firm{j_1})}$ let $(\worker{i_2},\firm{})$ be the
vertex containing $\firm{}$. As $(\worker{},\firm{})$ is not a blocking pair in $M_{(\worker{},\firm{j_1})}$ 
we note that $(\worker{i_1},\firm{})(\worker{},\firm{})(\worker{i_2},\firm{})$ is a path in $V\big(D(P)\big)$.
Therefore Property~($s$) holds, a contradiction.
\end{proof}

\begin{customlemma}{3}
Let $P$ be an instance of the matching problem and let $P^*$ denote the normal form of $P$.
The following two collection of pairs, $\match_{\Workers}$ and $\match_{\Firms}$, are both stable matchings in $P$.
\begin{center}
$\begin{array}{rcl} \vspace{0.2cm}
 \match_{\Workers} & = & \Set{\big(\worker{1}, \tau(\pref{\worker{1}^{*}})\big), \dots, \big(\worker{n}, \tau(\pref{\worker{n}^{*}})\big)} \\
 \match_{\Firms} & = & \Set{\big(\tau(\pref{\firm{1}^{*}}), \firm{1}\big), \dots, \big(\tau(\pref{\firm{n}^{*}}), \firm{n}\big)} \\
\end{array}$
\end{center}
\end{customlemma}

Before stating and proving Lemma~\ref{lemma:extremalMatches}, we introduce the following notation. Let $A_{\Workers}^*$ denote the arcs from $A_{\Workers}$ with both endpoints still in $D^*(P)$.
Analogously let $A_{\Firms}^*$ denote the arcs from $A_{\Firms}$ with both endpoints still in $D^*(P)$.
That is $A\big(D^*(P)\big)=A_{\Workers}^* \cup A_{\Firms}^*$.
Let $d^{*,+}_{\Workers}(x)$ denote the number of arcs out of $x \in V\big(D^*(P)\big)$
in $A_{\Workers}^*$ and analogously let $d^{*,+}_{\Firms}(x)$ denote the number of arcs out of $x \in V\big(D^*(P)\big)$
in $A_{\Firms}^*$.

\begin{customlemma}{3$'$} \label{lemma:extremalMatches}
Let $P$ be an instance of the matching problem. The following two sets, $M_{\Workers}$ and $M_{\Firms}$, are both stable matchings in $P$.
\begin{center}
$\begin{array}{rcl} \vspace{0.2cm}
 M_{\Workers} & = & \{ x \in V\big(D^*(P)\big) \; | \; d_{\Workers}^{*,+}(x)=0 \} \\
 M_{\Firms} & = & \{ x \in V\big(D^*(P)\big) \; | \; d_{\Firms}^{*,+}(x)=0 \} \\
\end{array}$
\end{center}
\end{customlemma}

\paragraph{Proof of Lemma~\ref{lemma:extremalMatches} (and hence of Lemma~\ref{lemma:extremalMatchesE}).}
\begin{proof}
By Lemma~\ref{lemma:iduaSameMatchingsE}, we note that there does exist some stable matching in $D^*(P)$.
So for every $\worker{} \in \Workers{}$ some vertex $(\worker{}, \firm{})$ belongs to $V(D^*(P))$.
Let $M_{\Workers}$ be defined as above and note that for every $\worker{} \in \Workers{}$ there exists exactly one vertex $(\worker{}, \firm{\worker{}})$ in 
$M_{\Workers}$. We first show that if $\worker{i}, \worker{j} \in \Workers{}$ are distinct then $\firm{\worker{i}}$ and $\firm{\worker{j}}$ are distinct.
For the sake of contradiction, assume that this is not the case, and $\firm{\worker{i}}=\firm{\worker{j}}$. Let $\firm{} = \firm{\worker{i}}=\firm{\worker{j}}$
and without loss of generality  assume that $(\worker{i}, \firm{}) (\worker{j}, \firm{}) \in A(D^*(P))$. However this implies that we could perform Reduction~$\reduce$ 
on $(\worker{j}, \firm{})$, which would imply that we would have deleted $(\worker{i}, \firm{})$, a contradiction.  Therefore  $\firm{\worker{i}}\not=\firm{\worker{j}}$ for all distinct
$\worker{i}$ and $\worker{j}$. Thus $M_{\Workers}$ is a matching.

By the proof of Lemma \ref{lemma:orderIndependentE}, every $(\worker{}, \firm{}) \in V\big(D^*(P)\big)$ either belongs to $M_{\Workers}$ or is an in-neighbour 
of $(\worker{}, \firm{\worker{}})$ which belongs $M_{\Workers}$.
Thus, there are no blocking pairs in $M_{\Workers}$, which implies that $M_{\Workers}$ is indeed a stable matching in $D^*(P)$.
By Lemma~\ref{lemma:iduaSameMatchingsE} $M_{\Workers}$ is therefore also a stable matching in $P$.

The fact that $M_{\Firms}$ is a stable matching in $P$ can be shown analogously.
\end{proof}


\begin{customthm}{1}
Let $P$ be an instance of the matching problem and let $P^{*}$ be its associated normal form. Then the following three statements are equivalent.
\begin{enumerate}[label=(\alph*)]
 \item $P$ has a unique stable matching.
 \item The normal form of $P$, $P^{*}$, is acyclic.
 \item In the normal form, $P^{*}$, every market participant's preference list a singleton.
\end{enumerate}

\end{customthm}

\begin{customthm}{1$'$}\label{thm:uniqueStable}
Let $P$ be an instance of the matching problem, and let $D(P)$ be the associated matching digraph, and let $D^*(P)$ be the matching digraph of the normal form. Then the following three statements are equivalent.

\begin{enumerate}[label=(\alph*)]
 \item $P$ has a unique stable matching.
 \item $D^*(P)$ contains no cycle.
 \item $D^*(P)$ consists of $n$ isolated vertices (that is, $|V(D^*(P))|=n$ and $|A(D^*(P))|=0$).
\end{enumerate}

\end{customthm}

The proof of Theorem~\ref{thm:uniqueStable} requires the following lemma, Lemma~\ref{lemma:emptyArc}.

\begin{customlemma}{4} \label{lemma:emptyArc}
Let $P$ be an instance of the matching problem.
If $A\big(D^*(P)\big)$ is nonempty, then $P$ contains at least two distinct stable matchings.
\end{customlemma}

\begin{proof}
Let $P$ be an instance that contains a stable matching, such that $A\big(D^*(P)\big) \not= \emptyset$.
Without loss of generality assume that $(\worker{},\firm{i}) (\worker{},\firm{j})$ is an arc in $A\big(D^*(P)\big)$.
Let $(\worker{},\firm{k}) \in V\big(D^*(P)\big)$ be chosen such that
$d_{\Workers}^{*,+}((\worker{},\firm{k}))=0$ ($j=k$ is possible) and note that $i \not= k$.
This implies that $(\worker{},\firm{k}) \in  M_{\Workers}$ (defined in Lemma~\ref{lemma:extremalMatches}).

If $M_{\Workers} \not= M_{\Firms}$ then $P$ contains at least two
distinct stable matchings (namely $M_{\Workers}$ and $M_{\Firms}$).
So, we may assume that $M_{\Workers} = M_{\Firms}.$
Hence, $(\worker{},\firm{k}) \in  M_{\Firms}$. However this implies that $(\worker{},\firm{i})$ will be deleted
by Reduction~$\reduce$, as $(\worker{},\firm{i})$ is an in-neighbour of $(\worker{},\firm{k})$ and there is no arc $a\in A_{\Firms{}}$ 
such that $(\worker{},\firm{k})$ is the tail of $a$. This is a contradiction to the fact that we cannot 
perform Reduction~$\reduce$ on $D^*(P)$.
\end{proof}

\paragraph{Proof of Theorem \ref{thm:uniqueStable} (and hence of Theorem \ref{thm:uniqueStableE}).}
\begin{proof}
We first prove that (a) and (b) are equivalent.
Assume that (b) holds.  If $(\worker{i},\firm{i})$ and 
$(\worker{j},\firm{j})$ are distinct vertices in $V\big(D^*(P)\big)$ then $\worker{i} \not= \worker{j}$ and $\firm{i} \not= \firm{j}$
as otherwise there would be an arc between the two vertices. So the $n$ vertices in $V\big(D^*(P)\big)$ form a 
matching of $P$.
This matching is a stable matching, by Lemma~\ref{lemma:iduaSameMatchings}, as there are no blocking pairs.
Therefore (b) implies (a).

Now assume that (a) holds. By Lemma~\ref{lemma:emptyArc} we note that $A\big(D^*(P)\big) = \emptyset$.
 If $|V(D^*(P))|<n$, then $P$ contains no stable matching (by Lemma~\ref{lemma:iduaSameMatchings}), a contradiction.
If $|V(D^*(P))|>n$, then $A\big(D^*(P)\big) \not= \emptyset$, another contradiction (as there must then be two 
vertices in $D^*(P)$ with the same worker).
So $|V(D^*(P))|=n$ and (b) holds.

Clearly (b) implies (c). We will show that (c) implies (b), so assume that (c) holds.
If $A(D^*(P)) = \emptyset$, then (a) holds by Lemma~\ref{lemma:emptyArc}, and therefore (b) also holds by the above.
So we may assume that $A(D^*(P)) \not= \emptyset$. 
Let $p_0p_1\ldots p_r$ be a path in $D^*(P)$ with maximum number of vertices and, as  $A(D^*(P)) \not= \emptyset$, we have $r \geq 1$.
If $d_{D^*(P)}^+(p_r) > 0$, then the out-neighbour of $p_r$ must lie in $V(P)$, due to the maximality of $r$, which 
gives us a cycle in $D^*(P)$, contradicting the fact that (c) holds.  So $d_{D^*(P)}^+(p_r) = 0$.
However, this implies that $p_{r-1}$ will be deleted
by Reduction~$\reduce$, as $p_{r-1}p_r$ is an arc and 
there is no arc with tail $p_r$ in either $A_{\Workers{}}$ or $A_{\Firms{}}$. This is a contradiction to the fact that we cannot
perform Reduction~$\reduce$ on $D^*(P)$, which completes the proof.

\end{proof}

\section{Equivalent instances}\label{app:Equivalent}

In this part of the appendix we sketch how an analyst can recover all stable matching problems for which a given matching is the unique stable matching.

We begin with a particular matching, $\match$.
From there, we show how to ``build from scratch'' the full set of matching problems satisfying Definition~\ref{def:instance} (i.e., $2n$ complete preference lists) with the property that the matching in question is the unique stable matching for each.\footnote{We note that matching problems with incomplete preferences for which $\match$ is the unique stable matching can also be recovered, but we omit the details.}

The above is performed by effectively running IDUA ``in reverse''.
Such a procedure works because any matching can be viewed as a collection of preference lists with only one participant per list.
One can then augment the one-participant preference lists by continually adding participants in such a way that the IDUA procedure would have deleted the additional entries.
While this does not yield a succinct condition on preferences like acyclicity (or any of the conditions provided by the papers listed in Footnote \ref{fn:sufficient}) and the procedure may be computationally intensive, it is exhaustive.
That is, given any matching one can construct the entire set of equivalent instances, where equivalence is defined as possessing the same unique stable matching.

Lemma \ref{lemma:iduaSameMatchingsE} ensures that if two instances of the matching problem possess the same normal form then they must possess the same set of stable matchings; but the reverse implication need not hold.
However, parts \ref{thmStatementUnique} and \ref{thmStatementSingleton} of Theorem \ref{thm:uniqueStableE} confirm that when an instance of the matching problem possesses a unique stable matching, the IDUA procedure collapses each participant's preference list to a singleton. And as such the reverse implication of Lemma \ref{lemma:iduaSameMatchingsE} does hold for instances with a unique stable matching.

The fact above is useful as it allows the analyst to reconstruct the full set of complete preference lists for which a given matching is the unique stable matching. Let us elaborate on this. Consider Figure \ref{fig:reductions} and observe what happened in going from the second to last panel to the last panel.  By the rules of the deletion procedure, the vertex $(3, 2)$ was deleted and hence so were the arcs that made up the path $(2,2)\to(3,2), (3, 2)\to(3,3)$ (which signify that $\firm{3} \pref{\worker{3}} \firm{2}$ and $\worker{3} \pref{\worker{2}} \worker{2}$). Note however, that the matching digraph in the final panel would have been the same if the path in the second to last panel had been the reverse, $(3,3)\to(3,2), (3, 2)\to(2,2)$ ($\firm{2} \pref{\worker{3}} \firm{3}$ and $\worker{2} \pref{\firm{2}} \worker{3}$), or even if the path had instead been from vertex $(2,2)$ to $(3,3)$ via the vertex $(2, 3)$ ($\firm{3} \pref{\worker{2}} \firm{2}$ and $\worker{3} \pref{\firm{3}} \worker{2}$).

This insight can be built upon. While $\reduce$ reduces a matching problem without altering the set of stable matchings, one can define an operation, call it $\reduce^{-1}$, that expands a stable matching subproblem in such a way that the set of stable matchings is not altered. {The inverse operation $\reduce^{-1}$ simply adds participants to preference lists in such a way that they would be deleted by an application of $\reduce$, though we note that $\reduce^{-1}$ is a correspondence and hence $\reduce^{-1}(D)$ denotes the set of all appropriate digraphs.} Just as $\reduce$ can be applied repeatedly, the analyst can repeatedly apply $\reduce^{-1}$ until all $2n$ preference lists are complete, at which point the analyst has generated an instance of the matching problem for which the identified matching is the unique stable matching.

We will illustrate this inverse procedure for a specific example. When there are $n$ workers and $n$ firms, the number of possible instances where all participants have complete preferences is $(n!)^{2n}$, so for reasons of space we consider an example with $n=2$. We let $\Workers = \set{\worker{1}, \worker{2}}$ and $\Firms = \set{\firm{1}, \firm{2}}$, and we consider the matching $\match^{*} = \set{(\worker{1}, \firm{1}),(\worker{2}, \firm{2})}$.
We then ask:\ what is the full set of instances for which $\match^{*}$ is the unique stable matching? {Let us now show how to reverse engineer the set of preferences for which $\match^{*}$ is the unique stable matching.}

When $n=2$, there are a total of $(2!)^{2\times2} = 16$ possible instances. Since the matching digraph for an instance with $n=2$ is a $2\times 2$ grid, and a minimum cycle has 4 vertices, there are only two possible instances that possess a cycle (one going ``clockwise'' and the other ``anti-clockwise''). This means that for $n=2$ there are 14 instances with a unique stable matching. Since there are only two possible matchings when $n=2$, we can be sure that there are exactly $7$ instances for which $\match^{*} = \set{(\worker{1}, \firm{1}),(\worker{2}, \firm{2})}$ is the unique stable matching. These are given in Figure \ref{fig:reverse} below, where each panel presents one such instance both in terms of preferences and using the digraph representation.
In every digraph the vertices that comprise the original matching, $\match^{*}$, are shaded grey while the vertices that are added back are hollow. The arcs represent preferences.

\begin{figure}[h!]
\begin{center}
\begin{tabular}{|c|c|c|c|} 
\hline
\tikzstyle{vertexDOT}=[scale=0.25,circle,draw,fill]
\tikzstyle{vertexY}=[circle,draw, top color=gray!5, bottom color=gray!30, minimum size=11pt, scale=0.3, inner sep=0.99pt]
\tikzstyle{vertexZ}=[circle,draw, top color=black!40, bottom color=black!80, minimum size=11pt, scale=0.3, inner sep=0.99pt]
\tikzstyle{vertexW}=[circle,dotted, draw, top color=gray!1, bottom color=gray!1, minimum size=11pt, scale=0.3, inner sep=0.99pt]
\begin{tikzpicture}[scale=0.6]
\node [scale=0.9] at (3,5.6) {\mbox{ }};
\node (x11) at (1,5) [vertexY] {$(1,1)$};
\node (x12) at (3,5) [vertexW] {$(1,2)$};
\node (x21) at (1,3) [vertexW] {$(2,1)$};
\node (x22) at (3,3) [vertexY] {$(2,2)$};

\draw [->,line width=0.03cm] (x22) -- (x12);
\draw [->,line width=0.03cm] (x12) -- (x11);
\draw [->,line width=0.03cm] (x22) -- (x21);
\draw [->,line width=0.03cm] (x21) -- (x11);

\end{tikzpicture} \hspace{.1in}  &

\tikzstyle{vertexDOT}=[scale=0.25,circle,draw,fill]
\tikzstyle{vertexY}=[circle,draw, top color=gray!5, bottom color=gray!30, minimum size=11pt, scale=0.3, inner sep=0.99pt]
\tikzstyle{vertexZ}=[circle,draw, top color=black!40, bottom color=black!80, minimum size=11pt, scale=0.3, inner sep=0.99pt]
\tikzstyle{vertexW}=[circle,dotted, draw, top color=gray!1, bottom color=gray!1, minimum size=11pt, scale=0.3, inner sep=0.99pt]
\begin{tikzpicture}[scale=0.6]
\node [scale=0.9] at (3,5.6) {\mbox{ }};
\node (x11) at (1,5) [vertexY] {$(1,1)$};
\node (x12) at (3,5) [vertexW] {$(1,2)$};
\node (x21) at (1,3) [vertexW] {$(2,1)$};
\node (x22) at (3,3) [vertexY] {$(2,2)$};

\draw [->,line width=0.03cm] (x22) -- (x12);
\draw [->,line width=0.03cm] (x12) -- (x11);
\draw [->,line width=0.03cm] (x21) -- (x22);
\draw [->,line width=0.03cm] (x21) -- (x11);

\end{tikzpicture} &

\tikzstyle{vertexDOT}=[scale=0.25,circle,draw,fill]
\tikzstyle{vertexY}=[circle,draw, top color=gray!5, bottom color=gray!30, minimum size=11pt, scale=0.3, inner sep=0.99pt]
\tikzstyle{vertexZ}=[circle,draw, top color=black!40, bottom color=black!80, minimum size=11pt, scale=0.3, inner sep=0.99pt]
\tikzstyle{vertexW}=[circle,dotted, draw, top color=gray!1, bottom color=gray!1, minimum size=11pt, scale=0.3, inner sep=0.99pt]
\begin{tikzpicture}[scale=0.6]
\node [scale=0.9] at (3,5.6) {\mbox{ }};
\node (x11) at (1,5) [vertexY] {$(1,1)$};
\node (x12) at (3,5) [vertexW] {$(1,2)$};
\node (x21) at (1,3) [vertexW] {$(2,1)$};
\node (x22) at (3,3) [vertexY] {$(2,2)$};

\draw [->,line width=0.03cm] (x12) -- (x22);
\draw [->,line width=0.03cm] (x12) -- (x11);
\draw [->,line width=0.03cm] (x21) -- (x11);
\draw [->,line width=0.03cm] (x22) -- (x21);

\end{tikzpicture} &

\tikzstyle{vertexDOT}=[scale=0.25,circle,draw,fill]
\tikzstyle{vertexY}=[circle,draw, top color=gray!5, bottom color=gray!30, minimum size=11pt, scale=0.3, inner sep=0.99pt]
\tikzstyle{vertexZ}=[circle,draw, top color=black!40, bottom color=black!80, minimum size=11pt, scale=0.3, inner sep=0.99pt]
\tikzstyle{vertexW}=[circle,dotted, draw, top color=gray!1, bottom color=gray!1, minimum size=11pt, scale=0.3, inner sep=0.99pt]
\begin{tikzpicture}[scale=0.6]
\node [scale=0.9] at (3,5.6) {\mbox{ }};
\node (x11) at (1,5) [vertexY] {$(1,1)$};
\node (x12) at (3,5) [vertexW] {$(1,2)$};
\node (x21) at (1,3) [vertexW] {$(2,1)$};
\node (x22) at (3,3) [vertexY] {$(2,2)$};

\draw [->,line width=0.03cm] (x12) -- (x22);
\draw [->,line width=0.03cm] (x12) -- (x11);
\draw [->,line width=0.03cm] (x21) -- (x22);
\draw [->,line width=0.03cm] (x21) -- (x11);

\end{tikzpicture}

\\

$\begin{aligned}
\worker{1}:& &\firm{1} \pref{\worker{1}} \firm{2}\\
\worker{2}:& & \firm{1} \pref{\worker{2}} \firm{2}\\
\firm{1}:& & \worker{1} \pref{\firm{1}} \worker{2}\\
\firm{2}:& & \worker{1} \pref{\firm{2}} \worker{2}
\end{aligned}$
 
&

$\begin{aligned}
\worker{1}:& &\firm{1} \pref{\worker{1}} \firm{2}\\
\worker{2}:& & \firm{2} \pref{\worker{2}} \firm{1}\\
\firm{1}:& & \worker{1} \pref{\firm{1}} \worker{2}\\
\firm{2}:& & \worker{1} \pref{\firm{2}} \worker{2}
\end{aligned}$

&

$\begin{aligned}
\worker{1}:& &\firm{1} \pref{\worker{1}} \firm{2}\\
\worker{2}:& & \firm{1} \pref{\worker{2}} \firm{2}\\
\firm{1}:& & \worker{1} \pref{\firm{1}} \worker{2}\\
\firm{2}:& & \worker{2} \pref{\firm{2}} \worker{1}
\end{aligned}$

&

$\begin{aligned}
\worker{1}:& &\firm{1} \pref{\worker{1}} \firm{2}\\
\worker{2}:& & \firm{2} \pref{\worker{2}} \firm{1}\\
\firm{1}:& & \worker{1} \pref{\firm{1}} \worker{2}\\
\firm{2}:& & \worker{2} \pref{\firm{2}} \worker{1}
\end{aligned}$

\\

\hline\hline

\tikzstyle{vertexDOT}=[scale=0.25,circle,draw,fill]
\tikzstyle{vertexY}=[circle,draw, top color=gray!5, bottom color=gray!30, minimum size=11pt, scale=0.3, inner sep=0.99pt]
\tikzstyle{vertexZ}=[circle,draw, top color=black!40, bottom color=black!80, minimum size=11pt, scale=0.3, inner sep=0.99pt]
\tikzstyle{vertexW}=[circle,dotted, draw, top color=gray!1, bottom color=gray!1, minimum size=11pt, scale=0.3, inner sep=0.99pt]
\begin{tikzpicture}[scale=0.6]
\node [scale=0.9] at (3,5.6) {\mbox{ }};
\node (x11) at (1,5) [vertexY] {$(1,1)$};
\node (x12) at (3,5) [vertexW] {$(1,2)$};
\node (x21) at (1,3) [vertexW] {$(2,1)$};
\node (x22) at (3,3) [vertexY] {$(2,2)$};

\draw [->,line width=0.03cm] (x12) -- (x22);
\draw [->,line width=0.03cm] (x12) -- (x11);
\draw [->,line width=0.03cm] (x21) -- (x22);
\draw [->,line width=0.03cm] (x11) -- (x21);

\end{tikzpicture} &

\tikzstyle{vertexDOT}=[scale=0.25,circle,draw,fill]
\tikzstyle{vertexY}=[circle,draw, top color=gray!5, bottom color=gray!30, minimum size=11pt, scale=0.3, inner sep=0.99pt]
\tikzstyle{vertexZ}=[circle,draw, top color=black!40, bottom color=black!80, minimum size=11pt, scale=0.3, inner sep=0.99pt]
\tikzstyle{vertexW}=[circle,dotted, draw, top color=gray!1, bottom color=gray!1, minimum size=11pt, scale=0.3, inner sep=0.99pt]
\begin{tikzpicture}[scale=0.6]
\node [scale=0.9] at (3,5.6) {\mbox{ }};
\node (x11) at (1,5) [vertexY] {$(1,1)$};
\node (x12) at (3,5) [vertexW] {$(1,2)$};
\node (x21) at (1,3) [vertexW] {$(2,1)$};
\node (x22) at (3,3) [vertexY] {$(2,2)$};

\draw [->,line width=0.03cm] (x12) -- (x22);
\draw [->,line width=0.03cm] (x11) -- (x12);
\draw [->,line width=0.03cm] (x11) -- (x21);
\draw [->,line width=0.03cm] (x21) -- (x22);

\end{tikzpicture}

&

\tikzstyle{vertexDOT}=[scale=0.25,circle,draw,fill]
\tikzstyle{vertexY}=[circle,draw, top color=gray!5, bottom color=gray!30, minimum size=11pt, scale=0.3, inner sep=0.99pt]
\tikzstyle{vertexZ}=[circle,draw, top color=black!40, bottom color=black!80, minimum size=11pt, scale=0.3, inner sep=0.99pt]
\tikzstyle{vertexW}=[circle,dotted, draw, top color=gray!1, bottom color=gray!1, minimum size=11pt, scale=0.3, inner sep=0.99pt]
\begin{tikzpicture}[scale=0.6]
\node [scale=0.9] at (3,5.6) {\mbox{ }};
\node (x11) at (1,5) [vertexY] {$(1,1)$};
\node (x12) at (3,5) [vertexW] {$(1,2)$};
\node (x21) at (1,3) [vertexW] {$(2,1)$};
\node (x22) at (3,3) [vertexY] {$(2,2)$};

\draw [->,line width=0.03cm] (x12) -- (x22);
\draw [->,line width=0.03cm] (x11) -- (x12);
\draw [->,line width=0.03cm] (x21) -- (x11);
\draw [->,line width=0.03cm] (x21) -- (x22);

\end{tikzpicture}

&

\\

$\begin{aligned}
\worker{1}:& &\firm{1} \pref{\worker{1}} \firm{2}\\
\worker{2}:& & \firm{2} \pref{\worker{2}} \firm{1}\\
\firm{1}:& & \worker{2} \pref{\firm{1}} \worker{1}\\
\firm{2}:& & \worker{2} \pref{\firm{2}} \worker{1}
\end{aligned}$

&

$\begin{aligned}
\worker{1}:& &\firm{2} \pref{\worker{1}} \firm{1}\\
\worker{2}:& & \firm{2} \pref{\worker{2}} \firm{1}\\
\firm{1}:& & \worker{2} \pref{\firm{1}} \worker{1}\\
\firm{2}:& & \worker{2} \pref{\firm{2}} \worker{1}
\end{aligned}$

&

$\begin{aligned}
\worker{1}:& &\firm{2} \pref{\worker{1}} \firm{1}\\
\worker{2}:& & \firm{2} \pref{\worker{2}} \firm{1}\\
\firm{1}:& & \worker{1} \pref{\firm{1}} \worker{2}\\
\firm{2}:& & \worker{2} \pref{\firm{2}} \worker{1}
\end{aligned}$

&

\\

\hline

\end{tabular}

\end{center}
\caption{The 7 instances for which $\match^{*}$ is the unique stable matching.}
\label{fig:reverse}
\end{figure}

The above illustrates how an analyst can ``build up from scratch'' the full set of equivalent instances of matching problems with a unique stable matching.
We note that while the described procedure is computationally intensive, it is exhaustive.
We conclude by pointing out that the procedure allows the identification of all instances for which the stated matching is the unique stable matching, and not merely those matching problems wherein every preference list is complete (which is what we have shown).


\clearpage

\bibliographystyle{plainnat}
\bibliography{uniqueStableReferences.bib}

\end{document}